\documentclass[preprint,onecolumn,nofootinbib]{revtex4}
\pdfoutput=1
\usepackage[colorlinks=true,linkcolor=blue,urlcolor=blue,filecolor=black,citecolor=red,pdfstartview=FitV,pdftitle={},pdfsubject={},pdfkeywords={},pdfpagemode=None,bookmarksopen=true]{hyperref}
\usepackage{graphicx}
\usepackage{amsmath}
\usepackage{amsfonts}
\usepackage{amssymb,ulem}
\usepackage{color,xcolor}%
\usepackage{CJK}
\usepackage{subfigure}
\usepackage{amsthm,amsmath,amssymb}
\usepackage{mathrsfs}

\usepackage{dcolumn}
\usepackage{float}
\usepackage{multirow}
\begin{document}
\title{Investigation of transport properties of graphene Dirac fluid by holographic two-current axion coupling model}
\author{Cai-E Liu $^{1}$}
\thanks{liucaie88@snnu.edu.cn}
\author{ShaoGuang Zhang$^{1}$}
\thanks{zhangsg@snnu.edu.cn, corresponding author}
\affiliation{$^1$~ School of Physics and Information Technology, Shaanxi Normal University, Xi'an 710119, PR China}
	
\begin{abstract}
Recently, there has been great interest in the phenomenon of severe violation of the Wiedemann-Franz law in graphene Dirac fluids around 75 K, due to the strong coupling relativistic plasma near the neutral point, where traditional perturbation theory fails. To explain this phenomenon, this article proposes a holographic dual two-current axion coupling model, describing the interaction between electrons and holes in graphene near the charge neutrality point (CNP) and revealing the related physical mechanism. The study shows that the holographic two-current model aligns with experimental results at $100\mu m^{-2}$,  and correctly predicts conductivity as temperature increases. Additionally, the article explores the behavior of $\beta+\gamma$ and its impact on conductivity and thermal conductivity. The results suggest a frictional effect between electrons and holes. Consequently, this study provides us with a clearer understanding of the properties of graphene Dirac fluids and further confirms the reliability of the holographic duality method.
\end{abstract}	

\maketitle
\tableofcontents	

\section{Introduction}
\label{intro}
 In recent years, the study of graphene Dirac fluids has been a hot topic in the field of condensed matter physics. Dirac fluids refer to the relativistic plasma-like state of graphene near charge neutrality, where massless electrons and holes collide at high speeds, giving rise to a series of remarkable phenomena. In 2016, Crossno et al.\cite{crossno2016observation} reported a serious violation of the Wiedemann-Franz law in graphene Dirac fluids near the particle-hole symmetry point, where the thermal power significantly increases and approaches the hydrodynamic limit, especially at T-75K. This violation of the Wiedemann-Franz law in graphene was unexpected and not predicted theoretically. To explain this anomaly, Pongsangangan et al. hypothesized in \cite{pongsangangan2022hydrodynamics} that the opposite motion of electrons and holes in an electric field suppresses the electrical conductivity due to friction, while the same direction motion in a temperature field leads to infinite thermal conductivity. This suggests a significant difference between the Dirac fluid phenomenon and the Fermi liquid theory.

 Nowadays the holographic generalization of fluid dynamics method \cite{foster2009slow} has been proposed as a holographic model for strongly coupled plasmas with two different conserved U(1) gauge currents, in order to describe the properties of graphene, which is consistent with existing experimental data. The double current coupling model has been discussed in several articles, as seen in Ref.\cite{alishahiha2012charged,amoretti2014thermo,amoretti2015analytic,banks2015thermoelectric,cheng2015thermoelectric,donos2015navier,donos2016dc,ling2016characterization,ling2016novel,ling2018holographic,ling2021instability}. In these models, one of the gauge fields is regarded as the true Maxwell field, while the other gauge field serves as an auxiliary field. The introduction of the auxiliary field is to obtain an insulating phase with a hard gap, while the transport properties only focus on one gauge field. For example, in Ref. \cite{ling2016novel,ling2018holographic}, a novel holographic insulator and holographic superconductor have been constructed. In addition, Ref. \cite{ling2021instability} introduced the coupling between these two gauge fields and studied superconducting instability. Yunseok Seo et al.\cite{seo2017holography} proposed that both U(1) gauge fields are Maxwell fields, with one representing electrons and the other representing holes. They found that the transport coefficients in the two current model match experimental data better than those in the single current model.

Therefore, we will use a coupled model with axion fields and two U(1) gauge fields to investigate the interaction between electrons and holes under varying temperatures and electric fields. We aim to explore to what extent the two-current axion model can simulate experimental results and subsequently investigate the impact of the parameters $\beta$ and $\gamma$ on the electrothermal conductivity. It is important to note that the geometric shape of the system significantly influences the outcomes, requiring careful analysis when comparing with experimental data on graphene.

The paper is organised as follows: In Sec.II the holographic two-current model is introduced, and in Sec.III the thermoelectric transport properties of the system are analysed and calculated. In Sec.IV, the total electrical and thermal conductivity are discussed. In Sec.V, the two-current model is compared with experimental data. In Section VI, we explore the impact of parameters on the transport properties of graphene. Finally, Section VII provides a summary and discussion of the current research results and outlines the future research potential in the field of graphene Dirac physi
\section{Holographic model}
\label{sec:model}
After the previous work on the  holographic model with gauge-axion coupling \cite{gouteraux2016effective,baggioli2017higher,li2019simple,liu2022alternating,zhong2022transverse}, we consider the following   two-current model in  4-dimensional spacetime with
a gauge-axion coupling term between the two $ U(1)$ gauge fields, denoted by $A_\mu,\,B_\mu$  respectively, and the axion fields, denoted by $\phi^I,I=x,y$:
\begin{subequations}
 \begin{eqnarray}\label{1}
\qquad
&&S=\int d^4x\sqrt{-g}(L_0+L_{c})\,,\\
&&L_0=R+6-\frac{Z_1}{4}F^2-\frac{Z_2}{4}G^2-V(\mathrm{Tr}[X])\,,\\
&&L_c=\beta_{1} Tr[XF^2]+\beta_{2}Tr[XG^2]+2 \gamma Tr[XFG]\,.
 \end{eqnarray}
\end{subequations}

Here, $S_{0}$ is the action of the usual two-current theory, with $\Lambda=-3$  the cosmological constant, $F=dA$ and $G=dB$  the strength of the two gauge fields respectively and$\alpha_{1}$,$\alpha_{2}$ are both constants. The axion fields $\phi^{I}$ are added by a derivative term
\begin{equation}
X_{\mu\nu}=\frac{1}{2}\partial_{\mu}\phi^I\partial_{\nu}\phi^I
\end{equation}
where $\mathrm{Tr}[X]=\frac{1}{2}\partial_{\mu}\phi^I\partial^{\mu}\phi^I$ is the trace of the matrix $X_{\mu\nu}$ and $V(\mathrm{Tr}[X])$ means a general function of $\mathrm{Tr}[X]$.
The gauge-axion coupled term $S_c$ represents a minimum coupling between the two gauge fields  $A_\mu,\,B_\mu$  and the axion fields $\phi^I$.
Explicitly, the trace in $S_c$ means the trace of the matrix product:
\begin{eqnarray}\label{4}
&&\mathrm{Tr}[XF^2]=X_{\mu\nu}F^{\nu\alpha}F_\alpha^{~~\mu}\qquad
\mathrm{Tr}[XG^2]=X_{\mu\nu}G^{\nu\alpha}G_\alpha^{~~\mu}\,,\nonumber \\
&&\mathrm{Tr}[XFG]=X_{\mu\nu}F^{\nu\alpha}G_\alpha^{~~\mu}\,,
\end{eqnarray}
and $J_{1},J_{2},\gamma$ are coupling constants.
The equations of motion for the theory Eq.\eqref{1} are given by Eq.\eqref{vari}.
\begin{subequations}
	\begin{eqnarray}\label{vari}
		&R_{\mu\nu}-\frac{R}{2}g_{\mu\nu}+(\Lambda+\frac{V}{2})g_{\mu\nu}-V_XX_{\mu\nu}-\frac{Z_1}{2}(F_{\mu\alpha}F_\nu^{~~\alpha}-\frac{F^2}{4}g_{\mu\nu})-\frac{Z_2}{2}(G_{\mu\alpha}G_\nu^{~~\alpha}-\frac{G^2}{4}g_{\mu\nu})\nonumber\\
		&\qquad+\frac{\beta_1}{4}((XFF+FFX+FXF)_{\mu\nu}+\frac{\mathrm{Tr}(XF^2)}{2}g_{\mu\nu})\nonumber\\
		&\qquad+\frac{\beta_2}{4}((XGG+GGX+GXG)_{\mu\nu}+\frac{\mathrm{Tr}(XG^2)}{2}g_{\mu\nu})\nonumber\\
		&\qquad+\frac{\gamma}{4}((XFG+XGF+FXG+FGX+GXF+GFX)_{\mu\nu}+\mathrm{Tr}(XFG)g_{\mu\nu})=0\,,\\
		&\nabla_\mu( Z_1F^{\mu\nu}+\frac{\beta_1}{2}(XF+FX)^{\mu\nu}+\frac{\gamma}{2}(XG+GX)^{\mu\nu})=0\,,\\
		&\nabla_\mu( Z_2G^{\mu\nu}+\frac{\beta_2}{2}(XG+GX)^{\mu\nu}+\frac{\gamma}{2}(XF+FX)^{\mu\nu})=0\,,\\
		&\nabla_\mu((\beta_1F^{\mu\alpha}F_\alpha^{~~\beta}+\beta
		_2G^{\mu\alpha}G_\alpha^{~~\beta}+\gamma F^{\mu\alpha}G_\alpha^{~~\beta}
		+\gamma G^{\mu\alpha}F_\alpha^{~~\beta})\partial_\beta\phi^I-4V_X\nabla^\mu\phi^I)=0\,.
	\end{eqnarray}
 \end{subequations}
 
where we write, for instance,$(XFG)_{\mu\nu}=X_{\mu\alpha}F^{\alpha\beta}G_{\beta\nu}$ for short.
We take the ansatz  as
\begin{align}
&ds^2=\frac{1}{u^2}(-f(u)dt^2+\frac{1}{f(u)}du^2+dx^2+dy^2)\label{6}\\
&A=A_{t}(u)dt\,,\qquad B=B_{t}(u)dt\,,\qquad \phi^I=k \delta^I_i x^i\,.
\end{align}
with $k$ the strength constant of the axion fields. The  backgrounding solution is as follows:
\begin{subequations}
	\begin{eqnarray}
		&&A_t=\mu_A-\frac{\rho_A}{Z_1}u\,,\qquad B_t=\mu_B-\frac{\rho_B}{Z_2}u\,,\\ &&f=1+\frac{Z_1\mu_A^2+Z_2\mu_B^2}{4u_h^2}u^4-\Big(\frac{u}{u_h}\Big)^3\Big(\frac{Z_1\mu_A^2u_h^2+Z_2\mu_B^2u_h^2+1}{4}\Big) \nonumber \\
		&&\qquad +\Big(\int_{u_h}^u \frac{V(k^2 x^2)}{2 x^4} \, dx \Big)\,,
\end{eqnarray}
\end{subequations}
$u_h$ stands for the horizon.
$\mu_A,\mu_B$ are respectively the chemical potential of the gauge field $A_\mu$ and $B_\mu$,and the associated charge density $\rho_A$ and $\rho_B$ are given by
\begin{align}
	\rho_A=Z_1\mu_A/u_h\,,\qquad \rho_B=Z_2\mu_B/u_h \,.
\end{align}
Finally,the Hawking temperature as well as the  entropy density can be obtained as
\begin{subequations}
	\begin{eqnarray}
		T&&=\frac{1}{4\pi u_h}(3-\frac{V(k^2u_h^2)}{2}-\frac{u_h^2(Z_{1}\mu_{A}^2+Z_{2}\mu_{B}^2)}{4})\,,\\
		s&&=4\pi/u_h^2\,.
	\end{eqnarray}
\end{subequations}

We should point out that the coupling term $L_c$
does not affect the background solution, which is different from
the previously considered coupling method  $F_{\mu\nu}G^{\mu\nu}$. Since
we  attempt to simulate the frictional interaction between electrons and holes that only manifests during transport,
and a coupling method that does not affect the background solution seems more reasonable.
\section{ Thermoelectric transport properties}
\label{sec:Therm}
In this section, we shall calculate the  DC thermoelectric conductivities of the dual system,
following "membrane paradigm" procedure proposed in \cite{donos2014novel} (for more details, see \cite{ling2017holographic,baggioli2017diffusivities,2014JHEP...11..081D,Blake2013UniversalRF,blake2015quantum}).
 We apply around the background a constant electric field $E_{x}$ and $e_{x}$, associated with the gauge field $A_\mu$
 and $B_\mu$ respectively, and a constant  temperature gradient $\zeta=-\frac{\nabla T}{T}$.
They will generate corresponding a charge current $J_{A}^{x}$ and $J_{B}^{x}$, as well as a heat current $Q^x$.
The thermoelectric transport properties in the dual field theory is captured by the generalized Ohm's law.
\begin{eqnarray}
\label{ohm}
\left(\begin{array}{c}
J_{A}^{x} \\
J_{B}^{x}\\
Q^{x}
\end{array}
\right)
=
\left(\begin{array}{ccc}
\sigma_{AA}& \sigma_{AB} & \alpha_{A} T\\
\sigma_{BA}& \sigma_{BB} & \alpha_{B} T\\
\bar{\alpha_{A}}T &\bar{\alpha_{B}}T & \bar{\kappa}T
\end{array} \right)
\left(\begin{array}{c}
E_{x}\\
e_{x}\\
\zeta
\end{array} \right)
\end{eqnarray}

Due to the rotational invariance on the $x-y$ plane, we only consider the transport behavior along the $x$ direction.
In order to find the conductivities for the background in question, one takes into account
small perturbations around the background solution obtained from Einstein equations of motion. The perturbations imply
\begin{subequations}
	\label{pert}
	\begin{eqnarray}
		&&\delta A_{x}=(-E_{x}+A_{t}(u)\zeta_{x})t+a_{x}(u) \,,\\
		&&\delta B_{x}=(-e_{x}+B_{t}(u)\zeta_{x})t+b_{x}(u)
		\,,\\
		&&\delta g_{tx}=\frac{1}{u^2}(-\zeta_{x}f(u)t+h_{tx}(u))\,,\\
		&&\delta g_{ux}=\frac{1}{u^2}h_{ux}(u)\,,\\
		&&\delta \phi_{x}=\chi(u)\,.
	\end{eqnarray}
\end{subequations}
The key point in the "membrane paradigm" method is to construct a radial-conserved current, which is equal to the dual system current.
For the ingoing condition, the conserved current taken at the horizon gives $J_{A}^{x},J_{B}^{x}$ or $Q^x$.
Then, using the expressions Eq.\eqref{ohm} and Eq.\eqref{pert} we find the explicit values of the transport coefficients as
\begin{subequations}
	\label{10}
	\begin{eqnarray}
		&&\sigma_{AA}=\frac{J_{A}^{x}}{E_{x}}=\frac{\beta_{1}}{4}k^2u_h^2+Z_{1}+\frac{1}{4k^2}\frac{\pi_{A}^2}{\Sigma}\,,\\
		&&\sigma_{BB}=\frac{J_{B}^x}{e_{x}}=\frac{\beta_{2}}{4}k^2{u_h}^2+Z_{2}+\frac{1}{4k^2}\frac{\pi_{B}^2}{\Sigma}\,,\\
		&&\sigma_{AB}=\sigma_{BA}=\frac{1}{4}k^2u_h^2\gamma+\frac{1}{4k^2}\frac{\pi_{A}\pi_{B}}{\Sigma}\,,\\
		&&T\bar{\kappa}=\frac{Q_{x}}{\zeta_{x}}=\frac{4f'(u_h)^2}{k^2u_h^2\Sigma}\,,\\
		&&T\alpha_{A}=\frac{J_{A}^{x}}{\zeta_{x}}=-\frac{\pi_{A}f'(u_h)}{k^2u_h\Sigma}\,,\\
		&&T\alpha_{B}=\frac{J_{B}^{x}}{\zeta_{x}}=-\frac{\pi_{B}f'(u_h)}{k^2u_h\Sigma},
	\end{eqnarray}
\end{subequations}
where the Onsager relation holds $\alpha_{i}=\bar{\alpha_{i}}$, and the following physical quantities are defined for ease of giving by
\begin{subequations}
	\begin{eqnarray}
		&&\Sigma=4V'(k^2u_h^2)\!-u_h^2(\!\beta_1\mu_A^2\!+\beta_2\mu_B^2\!+2\gamma\mu_A\mu_B) \,,\\
		&&\pi_{A}=(k^2u_h^2(\beta_1\mu_A+\gamma\mu_B)+4Z_1\mu_A)\,,\\
		&&\pi_{B}=(k^2u_h^2(\beta_2\mu_B+\gamma \mu_A)+4Z_2\mu_B)\,.
	\end{eqnarray}
\end{subequations}

\section{Discussion on Total Electric Thermal Conductivity}
\label{disc}
Transport coefficients of graphene have been experimentally measured and theoretically analyzed in various works, as reviewed in \cite{sarma2011electronic, basov2014colloquium}. Additionally, many papers have employed holographic methods, as seen in \cite{hartnoll2007theory, crossno2016observation}. In a recent paper by Seo et al. \cite{seo2017holography}, thermal conductivity was measured and analyzed using a two-current model with holographic methods. The two currents represent the electron and hole currents in the system, with the Fermi energy tuned to coincide with the Dirac point. Notably, our action in Eq.\eqref{1} accounts for the impact of the interaction between them on the thermal transport of graphene near the CNP.

\begin{figure} [ht]
\centering
\includegraphics[width=8cm]{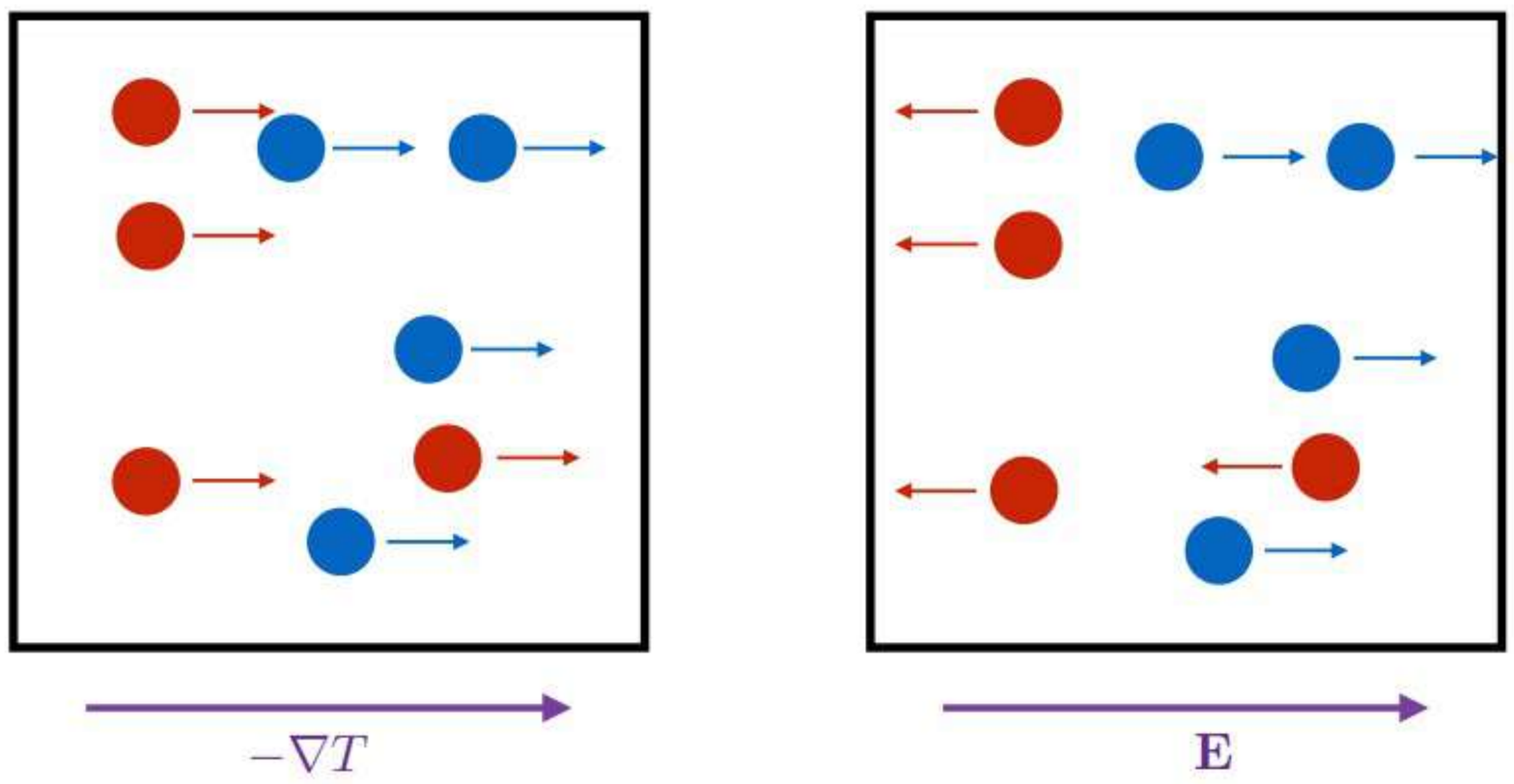}
\caption{\label{guess}The flow of electrons (red) vs. holes (blue) in the charge neutral Dirac fluid. Both electrons and holes move in the same direction in a temperature gradient, but move in opposite directions in an electric field\cite{lucas2018hydrodynamics}}
\end{figure}

One possible intuitive explanation for the violation of the Wiedemann-Franz law at the charge-neutral point in graphene at 75 K is discussed in \cite{lucas2018hydrodynamics}. When electrons and holes move in opposite directions in an electric field, their frictional interaction leads to a finite conductivity. On the other hand, when a temperature gradient is applied, electrons and holes move in the same direction, implying no friction and therefore infinite thermal conductivity in Fig.\ref{guess}.

To achieve an application and interpretation of Dirac fluids, we shall
first define the total electrical current, net charge density, and the
corresponding conductivities that can connect directly  with experimental quantities.
Using the electron-hole picture of Dirac fluids, while
gauge field $A_\mu$ and  $B_\mu$  will be explained as the
electromagnetic field from the  electrons and holes respectively.
Then $J_A^x,J_B^x$ represent the electron  and  hole, with $\rho_A=-en_{A},\rho_B=en_{B}$ the charge density.
In addition, the electric field $E_x$ and $e_x$ should be
identified $E_x=e_x$, for there is only one voltage driving
both the electron and hole current in practice.
We thus define the total electrical current, net charge density by
\begin{equation}
J^x=J^x_A+J^x_B\,,\qquad Q=\rho_A+\rho_B\,,
\end{equation}
Using the Ohm's law with $E_x=e_x$, the observed currents in Dirac fluids are given by
\begin{eqnarray}
J&=&J_{A}+J_{B}\\&=&(\sigma_{AA}+\sigma_{BB}+\!2\sigma_{AB})E_{x}+(\alpha_{A}+\alpha_{B})T\zeta_{x}\,,\\
Q&=&(\alpha_{A}+\alpha_{B})TE_{x}+\bar{\kappa}T\zeta_{x}\,,
\end{eqnarray}
and the experimental electrical conductivity $\sigma$ and thermal conductivity $\kappa$ can be extracted from this, as follows:
\begin{subequations}
\begin{eqnarray}
&&\sigma=\frac{J}{E_x}\Big|_{\zeta_{x}=0}=\sigma_{AA}+\sigma_{BB}+2\sigma_{AB}\,,\\
&&\kappa=\frac{Q}{\zeta_x}\Big|_{J=0}=\bar{\kappa}-\frac{(\alpha_{A}+\alpha_{B})^2T}{\sigma_{AA}+\sigma_{BB}+2\sigma_{AB}}\,.
\end{eqnarray}
\end{subequations}

The underlying theory behind the thermal conductivity expression is based on \cite{lucas2018hydrodynamics}, which suggests that both electrons and holes are driven by the same electric field. Additionally,
The total charge is $Q=Q_{A}+Q_{B}$, and the net charge is $Q_{n}=Q_{A}-Q_{B}$.There exists a proportional relationship between the total charge and net charge, which can be expressed as follows:
\begin{equation}
Q_{n}=g_{n}Q\,.
\end{equation}

Building on the above,we assume $\beta_1=\beta_2=\beta\neq\gamma,Z_1=Z_2=Z$ is substituted into the action. In this case, the expression for electrical conductivity can be obtained through Ohm's law and the ingoing boundary condition:
\begin{small}
\begin{subequations}
\begin{eqnarray}
&\sigma_{0}\!&=2Z+\frac{k^2}{2}u_h^2(\beta+\gamma)\,,\\
&\bar{\kappa}&\!=-\frac{128\pi^2TZ^2}{k^2u_h^2(Q^2u_h^4(\beta(1+\!g_n^2)+\gamma(1-g_n^2)) -8Z^2V')}\,,\\
&\sigma&\!=\sigma_{0}(1-\frac{Q^2u_h^2(\!4Z\!+\!u_h^2(\!\beta+\!\gamma\!))
}{k^2(u_h^4Q^2(\beta(1\!+\!g_n^2)+\!\gamma(\!1-\!g_n^2\!)\!)\!-\!8Z^2V')}\!)\,,\\
&\kappa&\!=\!-\frac{128\pi^2TZ^2}{Q^2u_h^4(g_n^2k^2u_h^2(\beta-\!\gamma)-4Z)-\!8k^2u_h^2Z^2V'}\,,
\end{eqnarray}
\end{subequations}
\end{small}
Further simplification can be achieved by replacing $s=4\pi/{u_h^2}$
\begin{small}
\begin{eqnarray}
&&\sigma_{0}=2(Z+\frac{k^2\pi(\beta+\gamma)}{s})\,,\\
&&\sigma=\!\sigma_{0}(1-\!\frac{\pi Q^2 s \sigma_{0}}{k^2(2\pi^2Q^2(\beta(\!1+\!g_n^2)+\!\gamma(\!1-\!g_n^2))-\!s^2Z^2}\!)\,,\\
&&\bar{\kappa}=\!-\frac{4\pi s^3T Z^2}{k^2(2(\!1+\!g_n^2)\beta\pi^2Q^2\!-\!s^2Z^2-\!2(g_n^2-\!1)\pi^2Q^2\gamma)}\,,\\
&&\kappa=-\frac{4\pi s^3 T Z^2}{2\pi Q^2(g_n^2k^2\pi(\beta-\gamma)-sZ)-k^2s^2Z^2}\,.
\end{eqnarray}
\end{small}

\section{Comparing with experiments}
Previous to this, we  obtained the expression for the total thermal conductivity of graphene Dirac fluid.In this section, we will mainly explore the anomalous behavior of graphene Dirac fluid at 75K, which manifests as a significant increase in thermal conductivity near the charge neutrality point and a serious violation of the Wiedemann-Franz law.

\begin{figure}[ht]
	\centering
	\subfigure{		\includegraphics[width=7.8cm,height=5.8cm]{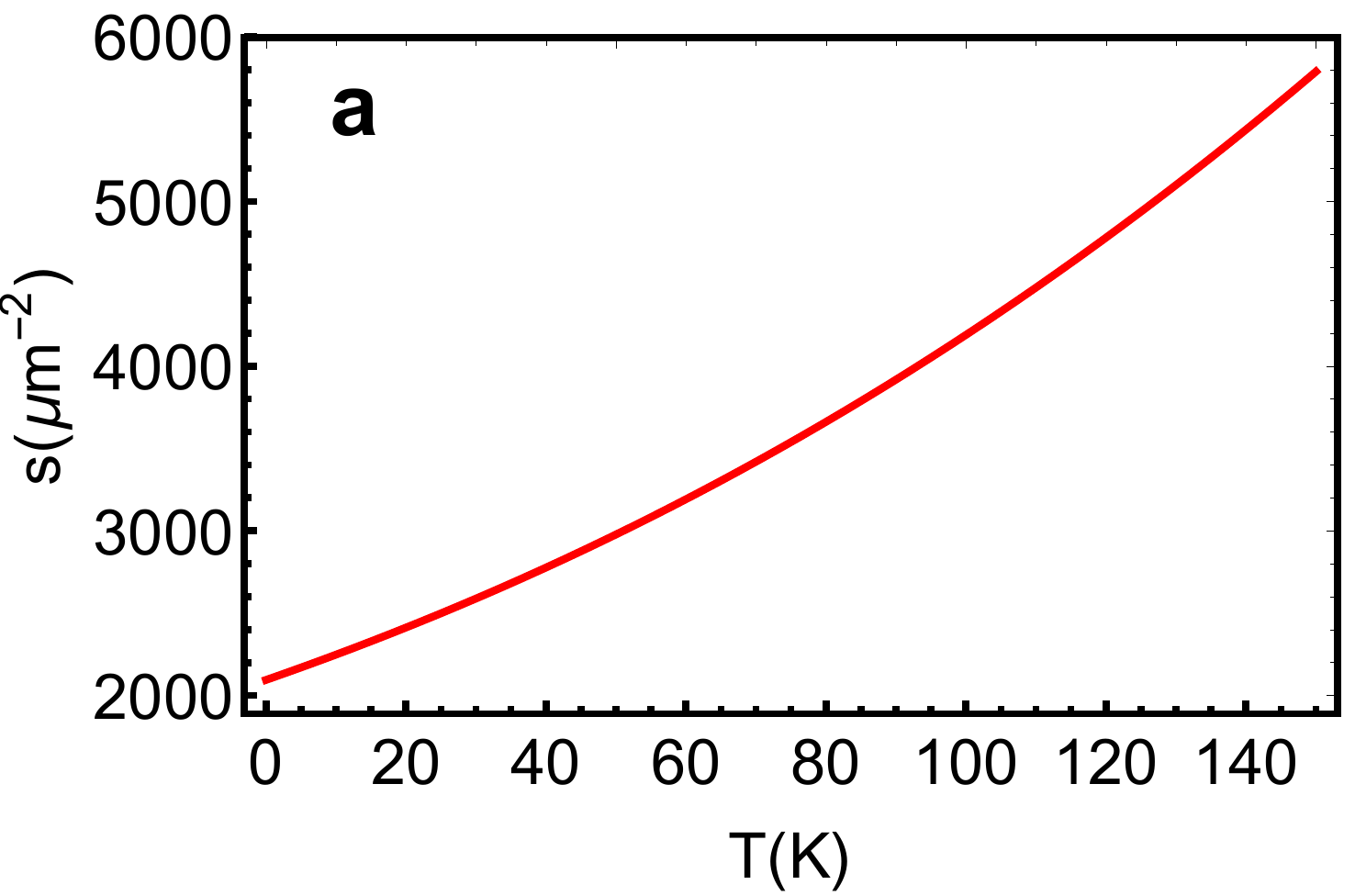}}
\subfigure{\includegraphics[width=7.8cm,height=5.8cm]{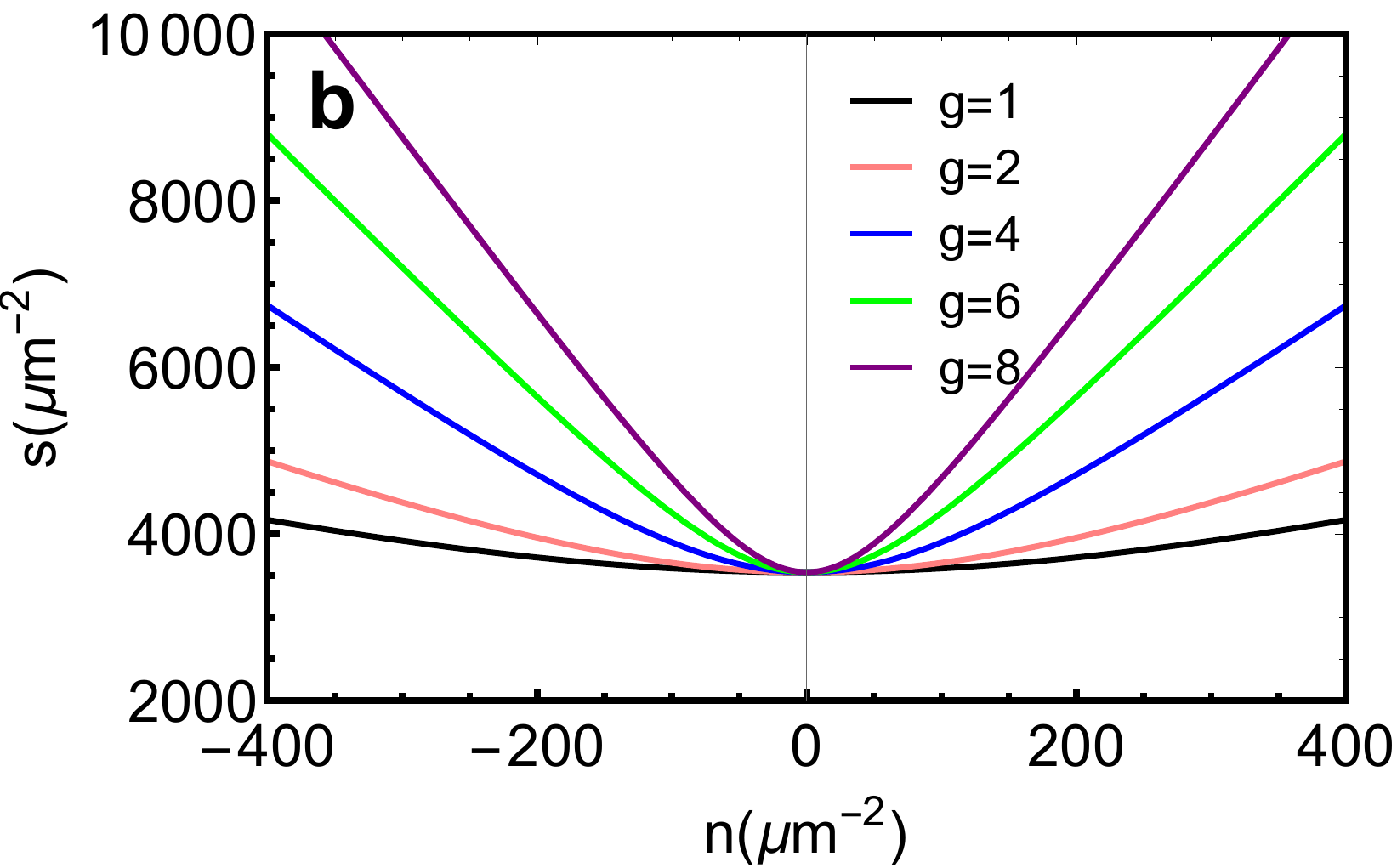}}
	 \caption{\label{s-T-n}(color online) (a) Entropy density $s$ as a function of temperature T at the charge neutrality point $(n=0)$. (b) Entropy density $s$ as a function of  charge density $n$ at $T=75K$.The plot shows the regulatory effect of charge fitting value $g$ on s,and as g increases, the rate of change of s with n becomes faster.}
\end{figure}

Before comparing with experiments, it was found that entropy density $s$ not only depends on $T$ but also varies with $n$. Since the transport coefficients of graphene are closely related to $s$, caution is required in the selection and handling of entropy density. As illustrated in Fig.\ref{s-T-n}, larger values of $g_{n}$ result in a more apparent change of entropy density concerning number density. To simplify the analysis, we select the case $v_{F}=6*10^6m/s,g_{n}=2$, where the change in entropy density is relatively small. In this case, we approximate the entropy density as a constant, denoted by $s=3538.33\mu m^{-2}$.

By observing the experimental data\cite{crossno2016observation}, we have obtained the values of $\sigma$ and $\kappa$ at CNP, which are presented below $T=75K$:
\begin{subequations}
\label{cnpalpha}
\begin{eqnarray}
 &&\kappa_{0}=\frac{4\pi sT k_{B}^2}{k^2\hbar}=7.7 nW/k\,, \\
 \label{sigma0}
 &&\sigma_{0}=2\frac{e^2}{\hbar}(Z+\frac{k^2\pi(\beta+\gamma)}{s})=0.338 k\Omega^{-1}\,.
\end{eqnarray}
\end{subequations}

In parameter selection, $Z$ is set to 1 based on \cite{li2022transport,gouteraux2016effective,baggioli2017higher,li2019simple,zhong2022transverse,iqbal2009universality,donos2014thermoelectric,kim2015thermoelectric} practices for consistency and comparability. Subsequently, $k^2$ is determined as $1000\mu m^{-2}$ using the temperature formula, and $s=3538.33\mu m^{-2}$. These key parameters reveal a constant difference between $\beta$ and $\gamma$ at a fixed temperature.Solving the conductivity equation yields $\sigma_{0}=0.338 k\Omega^{-1}= 2e^2(1+\pi k^2(\beta+\gamma)/s)/\hbar$, resulting in $\beta+\gamma=-0.342$. In the thermal conductivity fitting process, it is established that $\gamma\geq0$. With other parameters fixed, careful adjustments to $\gamma$ and $\beta$ led to the optimal fit: $\gamma=8$ and $\beta=-8.342$. This systematic approach ensures rational and reliable parameter selection for a more accurate reflection of experimental observations.

In this appendix and in every subsequent appendix \ref{A2}, we will work in units where $\hbar=k_{B}=V_{F}=e=1$. It
is straightforward using dimensional analysis to restore these units. Following the previous settings, the relationship between the electrical and thermal conductivity and the charge density is obtained as follows:
\begin{small}
\begin{subequations}
\label{skapn}
\begin{eqnarray}
&&\kappa(n)\!=\frac{\kappa_{0}}{1-\frac{2\pi(g_n^2k^2\pi(\beta-\gamma)-sZ)n^2}{k^2s^2Z^2}}\,,\\
&&\sigma(n)\!=\sigma_{0}(\!1-\!\frac{n^2}{\frac{2\pi n^2k^2}{s\sigma_{0}}(\beta(\!1+\!g_n^2)+\!\gamma(\!1-\!g_n^2))\!-\!\frac{k^2 s Z^2}{\pi\sigma_{0}}})\,.
\end{eqnarray}
\end{subequations}
\end{small}
\begin{figure}[ht]
	\centering
\subfigure{\includegraphics[width=7.6cm,height=6cm]{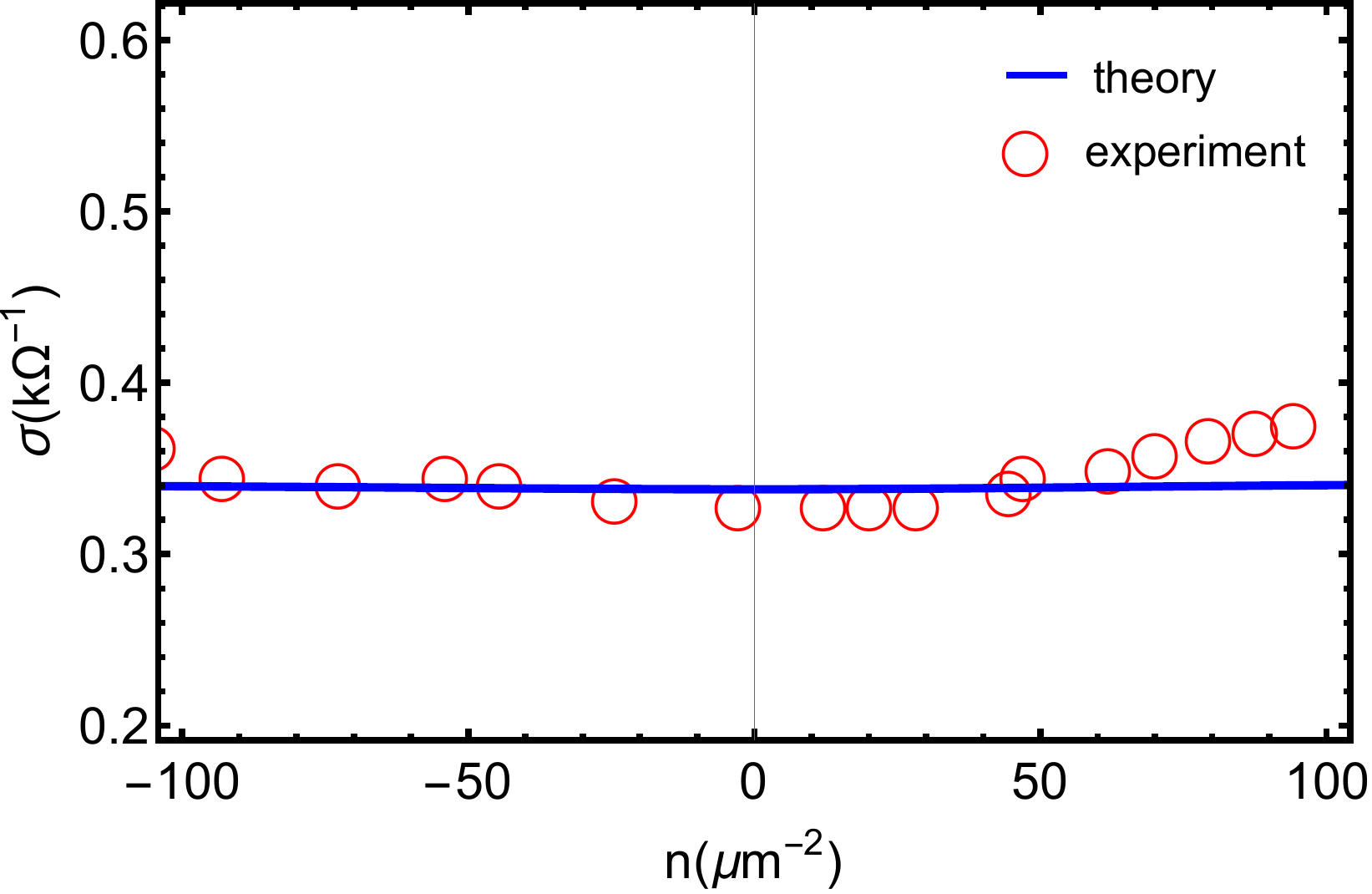}}
\subfigure{\includegraphics[width=7.6cm,height=6cm]{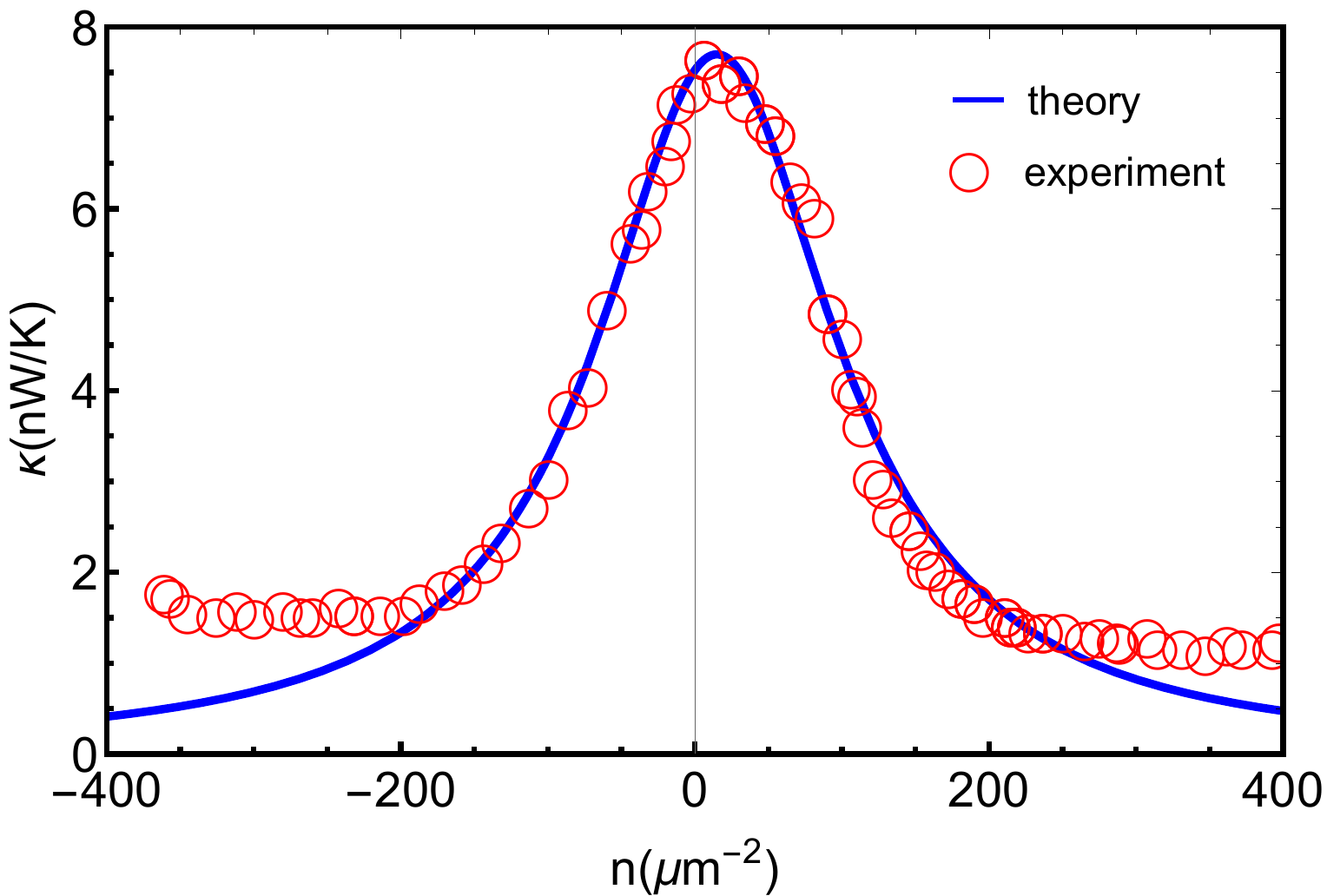}}
	\caption{ \label{exper-1}Theory vs. Data: (left panel) The measured conductivity $\sigma$ as functions of charge density n. (right panel) The measured thermal conductivity $\kappa$ as functions of charge density n. Blue lines is the analytic Eq.\eqref{skapn} and the red dot are for experimentally data used in\cite{crossno2016observation,lucas2016transport}}
\end{figure}
Our model demonstrates strong agreement with experimental data across the range of $-100 \mu m^{-2} \leq n \leq 100 \mu m^{-2}$, as depicted in Fig.\ref{exper-1}. In the absence of disorder, conductivity remains finite due to the establishment of electron-hole equilibrium at specific temperatures. This equilibrium nullifies the total momentum induced by an electric field, obviating the need for contaminants to relax momentum. Instead, electrons and holes are bound together through the drag mechanism \cite{pongsangangan2022hydrodynamics}. Moreover, in Eq.\eqref{cnpalpha}, it is evident that $\beta+\gamma$ exclusively influences the electrical conductivity $\sigma_{0}$, while the thermal conductivity $\kappa_{0}$ is solely dependent on entropy density and temperature. We hypothesize that at the charge neutrality point (CNP), $\beta$ and $\gamma$ exhibit behavior reminiscent of the Drag effect.

\subsection{Discussion on $\sigma(T)$and $\kappa(T)$}
\label{subsec:1}
In this section, we will concentrate on examining the characteristics of thermoelectric transport coefficients at various temperatures. Through experimental observations, it is noted that, at zero charge density, the conductivity $\sigma_{0}$ exhibits a monotonic increase with temperature. This observed behavior aligns with our intuitive understanding of electrical conductivity in everyday scenarios.

Subsequently, substituting $s(T)$ into the expressions for the conductivity $\sigma_{0}(T)$ and thermal conductivity $\kappa_{0}(T)$ at the neutrality point.
\begin{subequations}
\label{skpT}
\begin{eqnarray}
&&\sigma_{0}(T,n)|_{n=0}=2(Z+\frac{k^2\pi(\beta+\gamma)}{s(T)})\,,\\
&&\kappa_{0}(T,n)|_{n=0}=\frac{4\pi s(T) T}{k^2}\,.
\end{eqnarray}
\end{subequations}

\begin{figure}[ht]
	\centering
\subfigure{\includegraphics[width=7.6cm,height=6cm]{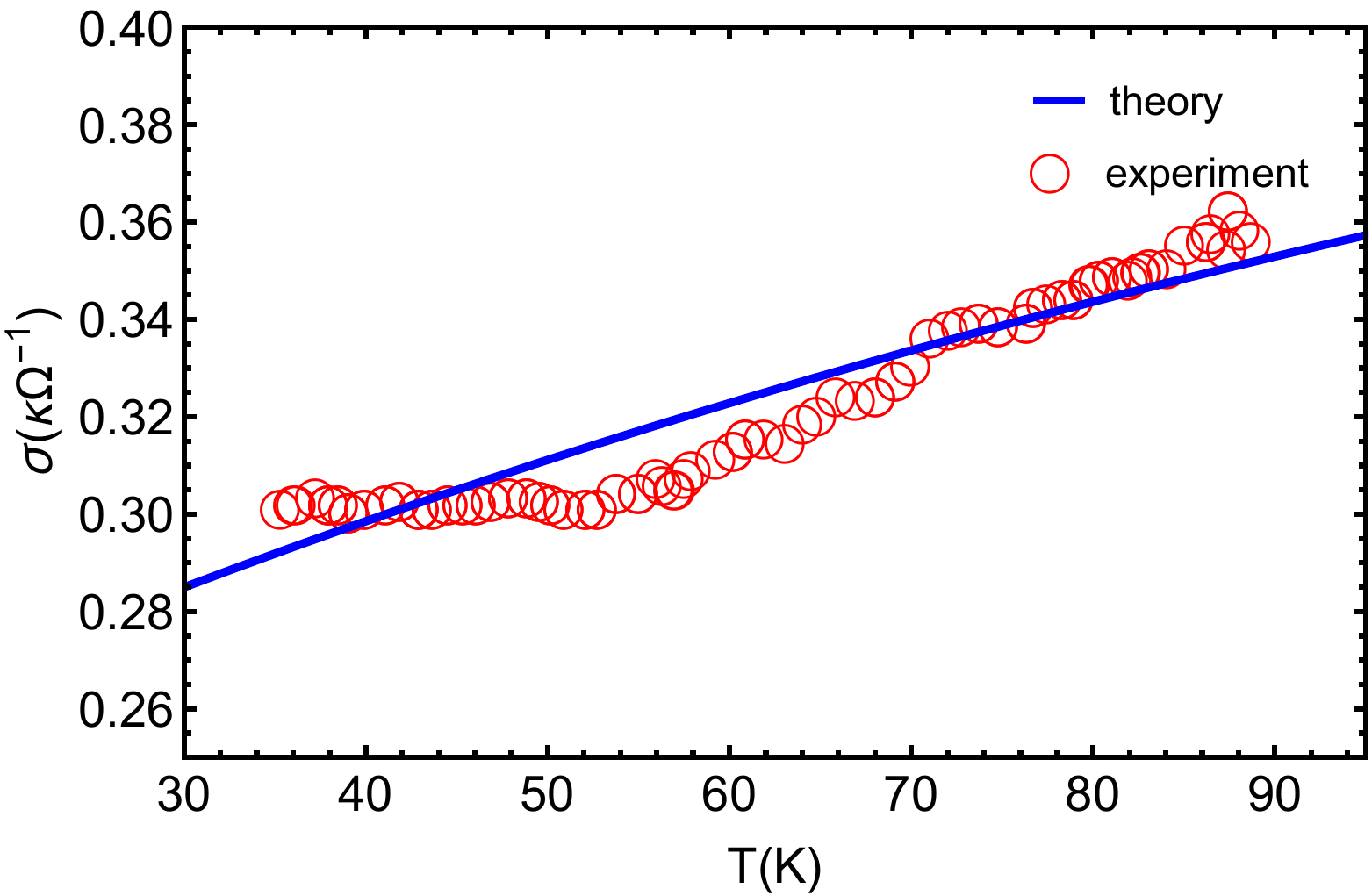}}
\subfigure{\includegraphics[width=7.6cm,height=5.9cm]{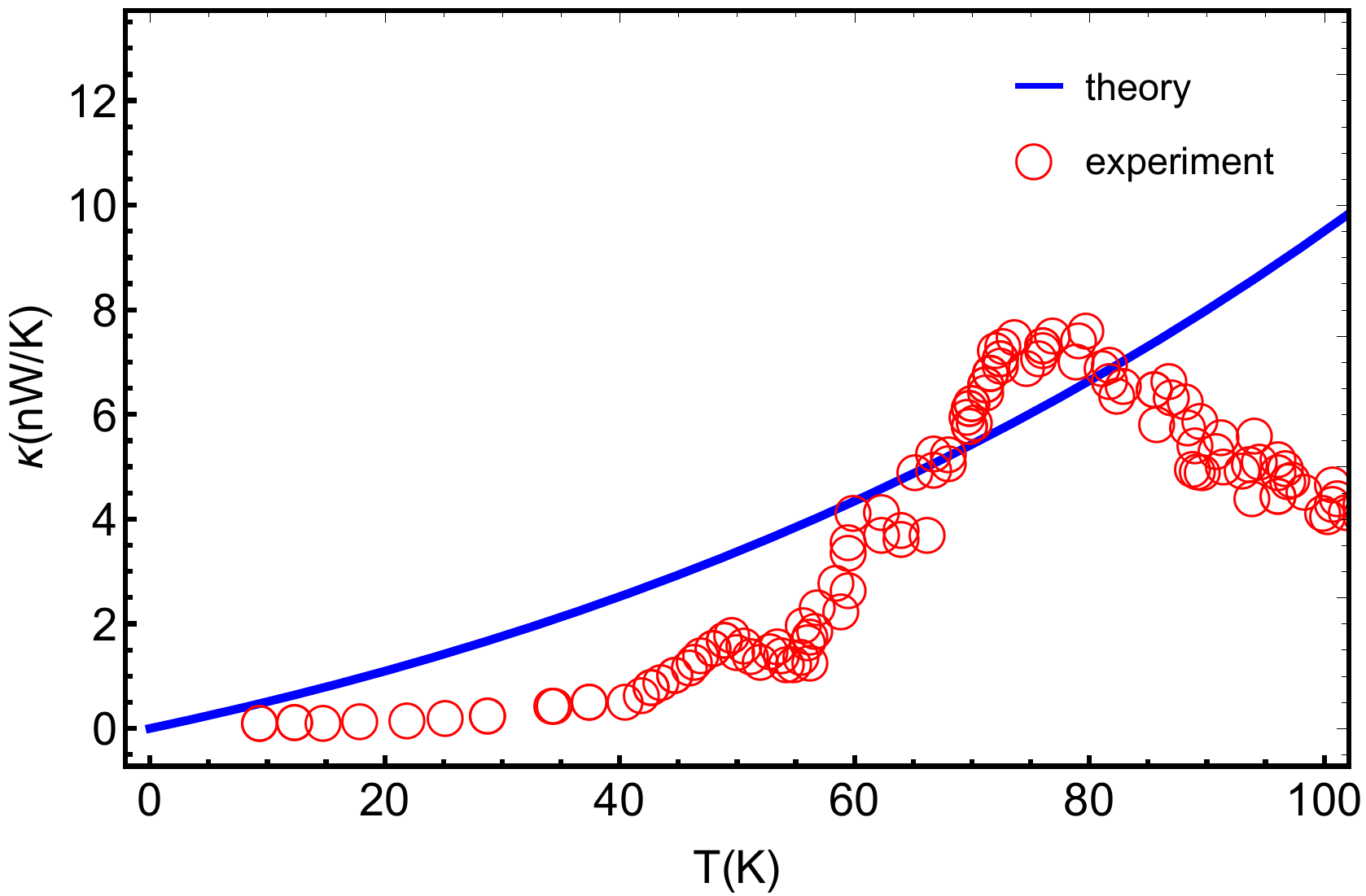}}
	\caption{\label{tem-s-k}Theory vs. Data:(left panel) The measured conductivity $\sigma_{0}$ as functions of temperature T at the charge neutrality point $(n=0)$. (right panel) The measured thermal conductivity $\kappa_{0}$ as functions of temperature T at the charge neutrality point $(n=0)$.Blue lines is the analytic Eq.\eqref{skpT} and the red dot are for experimentally data used in\cite{crossno2016observation,lucas2016transport}}
\end{figure}

Our theory not only provides quantitative fits for $\kappa_{0}(T)$ and $\sigma_{0}(T)$ but also faithfully reproduces existing research results as much as possible, which is shown in Fig.\ref{tem-s-k}. Particularly noteworthy is that our theory captures the phenomenon of increasing $\sigma_{0}(T)$ with rising temperature, offering valuable insights into the momentum relaxation mechanism.

 In comparison to the work by Yunseok Seo et al.\cite{seo2017holography,song2020determination,narozhny2021hydrodynamic}, $\sigma_{0}(T)$ is no longer a constant; we introduced an additional term $\pi k^2(\beta+\gamma)/s(T)$. The inclusion of this term enhances the precision of our simulation results, providing a more accurate reflection of the trend in electrical conductivity with increasing temperature. However, the correspondence between $\kappa_{0}(T)$ and experimental results is not ideal, and we speculate that this discrepancy may be due to the temperature-dependent relationship of entropy density.

\subsection{The Wiedemann-Franz ratio}
\label{WF}

Fermi liquid has a noteworthy property\cite{ziman2001electrons}: the Lorenz ratio of thermal conductivity to electric conductivity remains constant at low temperatures. This property is known as the Wiedemann-Franz (WF) law\cite{mermin1978solid}:
\begin{equation}
L=\frac{\kappa_{e}}{\sigma T}=\frac{\pi^2}{3}\frac{(k_{B})^2}{e^2}=L_{0}
\end{equation}

However, the WF law is fragile.
This law has been found to be broken in the strongly interacting non-Fermi liquids\cite{mahajan2013non,kim2009violation}.
Because of the inelastic scattering between charged and neutral degrees of freedom, it can be ascribed to heat and charge transport in different ways \cite{mahajan2013non}.

For the charge neutral point region of graphene's Dirac fluid, it has been experimentally observed that the WF law is strongly violated at around 75K.Therefore, we obtained expressions for the Lorenz number $L(n)$ based on the transport coefficient of our model.
\begin{equation}\label{L}
 L(n)|_{T=75K}=\frac{\kappa_{0}(T)}{(1-\frac{2\pi(g_n^2k^2\pi(Z-\gamma)-s(T)w)n^2}{k^2s(T)^2w^2})\Big
(\sigma_{0}(T)(1-\frac{n^2}{\frac{2\pi n^2k^2}{s(T)\sigma_{0}(T)}(Z(1+g_n^2)+\gamma(1-g_n^2))-\frac{k^2 s(T) w^2}{\pi\sigma_{0}(T)}})\Big)T}
\end{equation}
\begin{figure}[ht]
	\centering
\subfigure{\includegraphics[width=7.6cm,height=6cm]{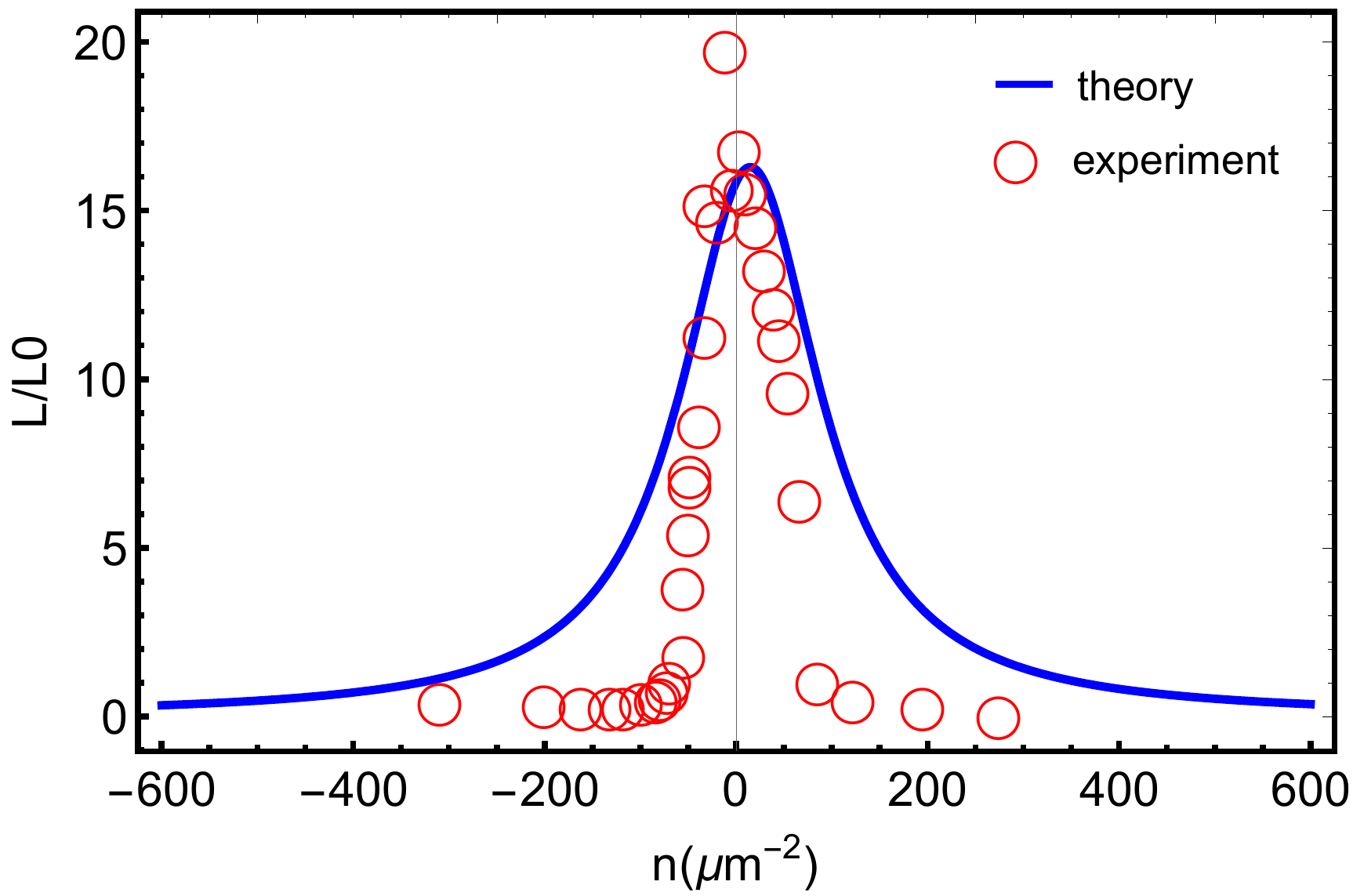}}
\subfigure{\includegraphics[width=7.6cm,height=5.9cm]{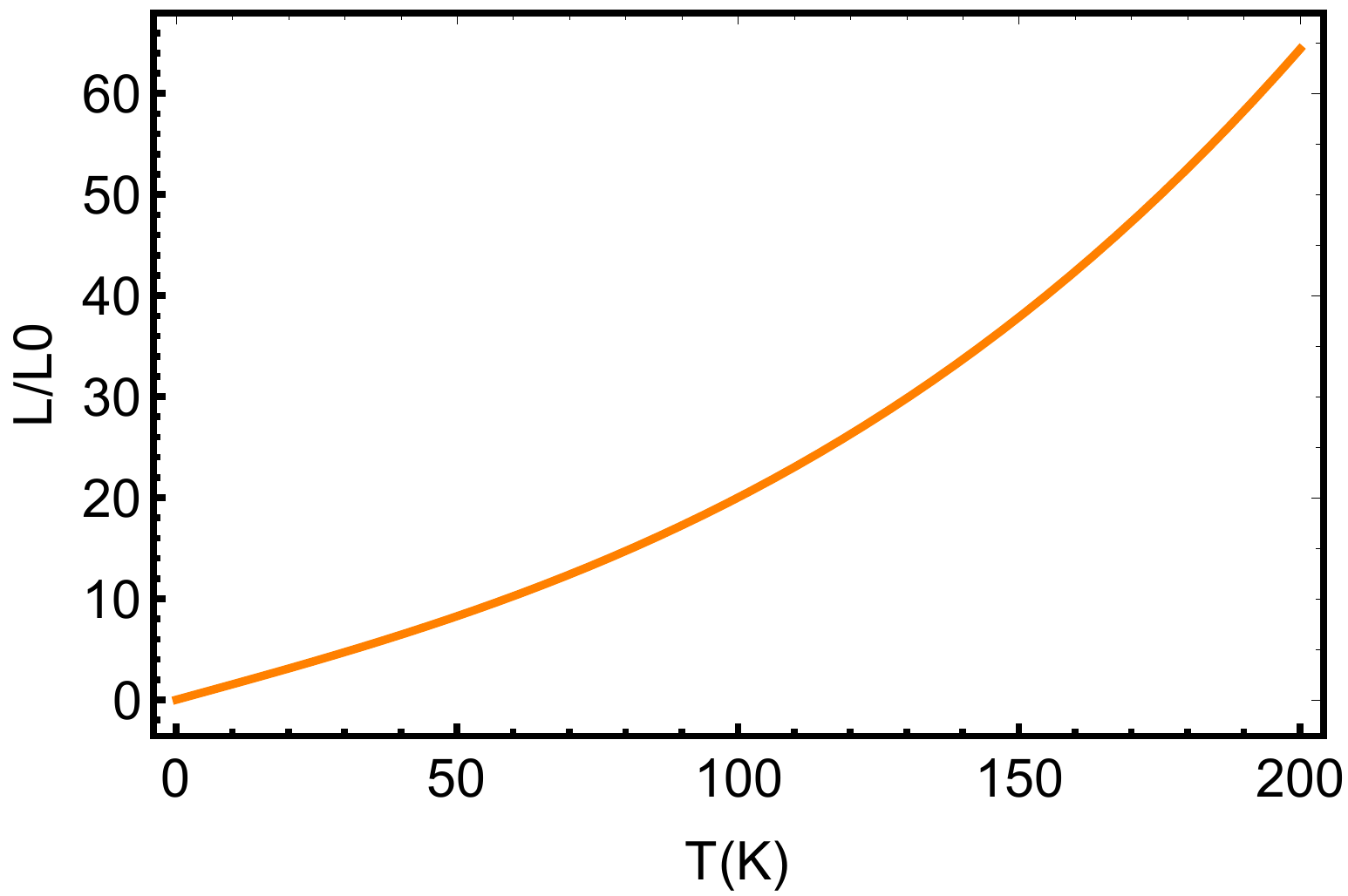}}
  \caption{\label{Fig.3}(left panel) The measured Lorenz number $L/L0$ at T =60 K as function of charge density n.The red solid circle in the figure represents the experimental data results at 60K \cite{crossno2016observation},the blue solid line is the image taken by our model at 60K. (right panel) The measured Lorenz number as function of temperatures T,and the Lorentz number $L/L0$ here is represented by substituting $s (T) $into the Eq.$\eqref{L}$.}          
\end{figure}

 According to Fig.\ref{Fig.3}, we can see a significant peak of the Lorentz ratio near the charge neutral point, while it approaches zero in other regions. This indicates that the Lorentz ratio in graphene is no longer constant, accurately reflecting changes in the Dirac fluid state near the neutral point. Our model has effectively captured this characteristic. To put it simply, the unusual behavior of graphene near the neutral point is due to the uniform movement of electrons and holes in the same direction in the temperature field, without any frictional effects or loss of momentum. The lack of frictional effects and momentum loss significantly enhances the thermal conductivity of graphene, and the impact on its conductivity can be almost negligible.
 
 At high temperatures (above 100K), experimental data shows that the heat transfer from electrons to phonons is equivalent to the degree of electron diffusion\cite{fong2012ultrasensitive,fong2013measurement,crossno2015development}. However, this electron-phonon scattering leads to the degradation of Dirac fluids, resulting in a decrease in the Lorentz number $L$ at a temperature of approximately 100K. Our theory has shortcomings at this point, partly because our system did not take into account the scattering of phonons and electrons. On the other hand, our theoretical entropy density is linearly related to temperature, as shown in Fig.\ref{s-T-n}.
 
\subsection{Calculating the Seeback Coefficients}
This section will explore thermoelectric coefficient $\alpha$ properties of graphene and the Seebeck coefficient $S$. The distinctive feature of the Seebeck coefficient is its zero value at the neutral point, as electrons and holes of equal concentration move in the same direction in the temperature gradient field. Consequently, the current $J=0$, leading to $\alpha=0$.
 
 Simplifying according to Eq. \eqref{10}, we obtain the thermoelectric coefficients $\alpha_{A}$ and $\alpha_{B}$ separately:
\begin{eqnarray}
 &T\alpha_{A}=\frac{2\pi Q s T Z((1+g_{n})\beta k^2\pi+(1+g_{n})s Z+(1-g_{n})k^2\pi \gamma))}{k^2(2(1+g_{n}^2)\beta\pi^2Q^2-s^2Z^2+2(1-g_{n}^2)\pi^2Q^2\gamma)}\nonumber\\  &T\alpha_{B}=\frac{2\pi Q sTZ((1-g_{n})\beta k^2\pi+(1-g_{n})sZ+(1+g_{n})k^2\pi\gamma)}{k^2(2(1+g_{n}^2)\beta\pi^2Q^2-s^2Z^2+2(1-g_{n}^2)\pi^2Q^2\gamma)}
\end{eqnarray}

The definition for the total thermoelectric coefficient $\alpha$ and Seebeck coefficient $S$ are given as:
\begin{equation}
\label{T-alpha-se}
\qquad\qquad \alpha=\alpha_{A}+\alpha_{B}\qquad S=\frac{\alpha}{\sigma}
\end{equation}
Using Eq.\eqref{T-alpha-se} and Eq.\eqref{sigma0}, the expressions for the $\alpha$ and $S$ are derived as follows:
\begin{eqnarray}
    &&\alpha=\frac{k_{B}|e|}{\hbar}\frac{4\pi s Q Z( \pi k^2 (\beta+\gamma)+sZ)}{k^2\Theta}\\
    &&S=\frac{k_{B}}{|e|}\frac{\kappa_{0}Z Q}{2T\Theta(1-\frac{\pi Q^2 s\sigma_{0}}{k^2\Theta})}\\
    &&\Theta=2\pi^{2}Q^{2}(\beta(1+g_{n}^{2})+\gamma(1-g_{n}^2))-s^{2}Z^{2}
\end{eqnarray} 

\begin{figure}[ht]
  \centering
\subfigure{\includegraphics[width=7.6cm,height=6cm]{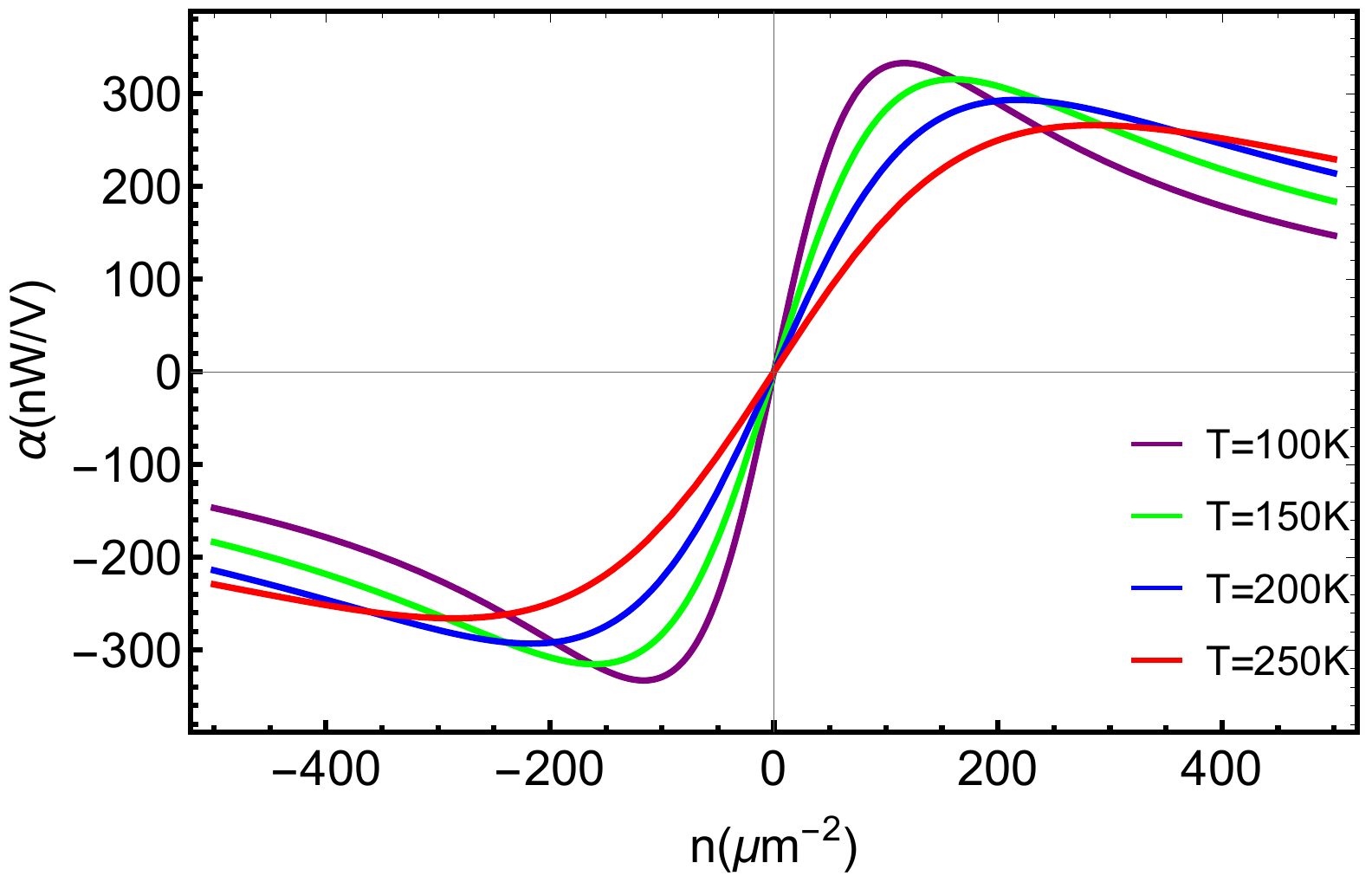}}
	\hspace{0.1in}
\subfigure{\includegraphics[width=7.6cm,height=5.9cm]{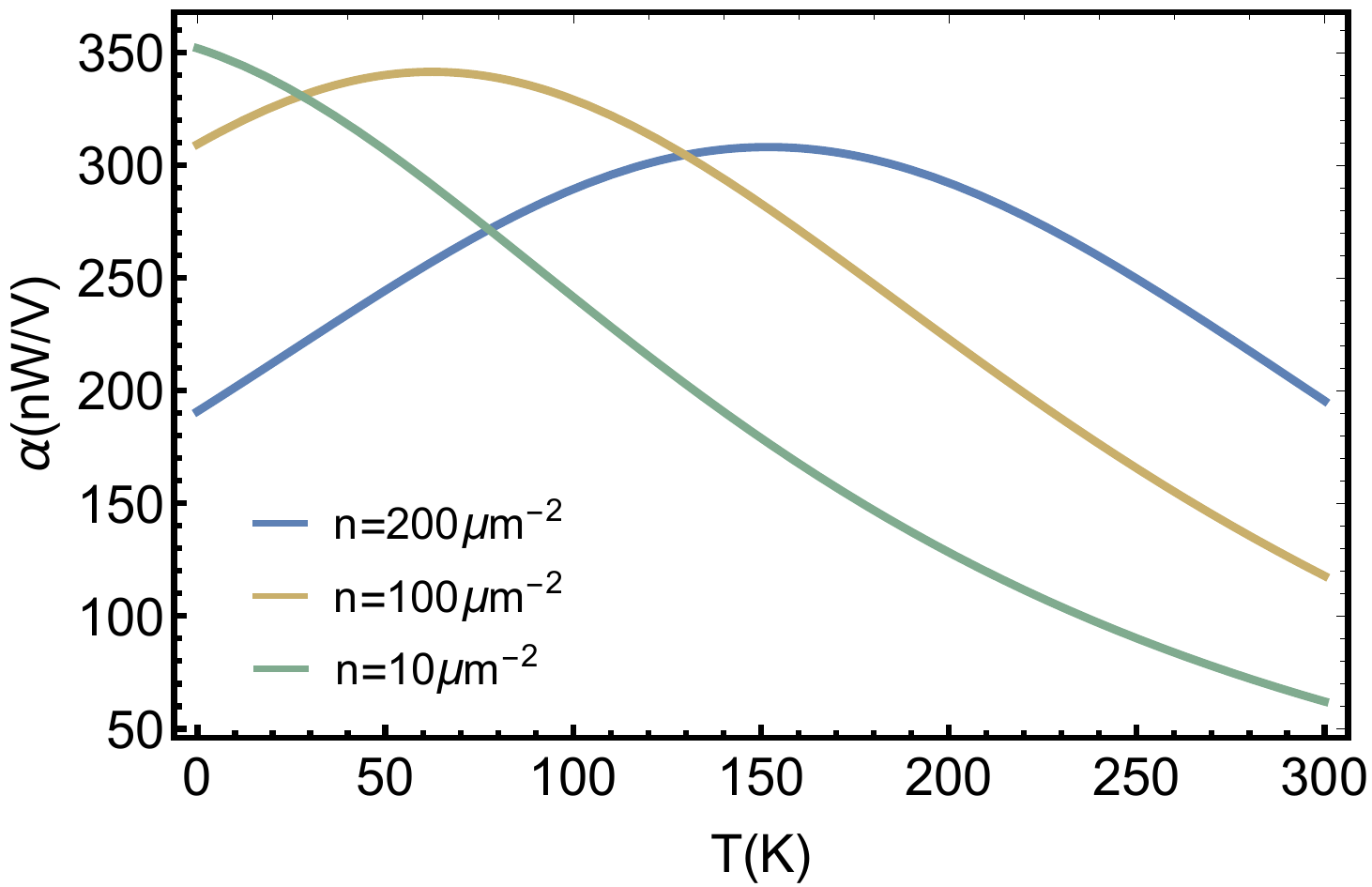}}
  \caption{\label{alpha-T-n}(color online) (left panel) Measured thermoelectric coefficient $\alpha$ as a function of density n for various temperatures. As we increase T, the maximum value of $\alpha$ shifts towards high concentration areas. (right panel) The measured thermoelectric coefficient $\alpha$ as a function of temperatures T for various density n. As we increase charge density n, the peak value of $\alpha$ will shift towards the high temperature region.}
  \end{figure}

\begin{figure}[ht]
  \centering
\subfigure{\includegraphics[width=7.6cm,height=6cm]{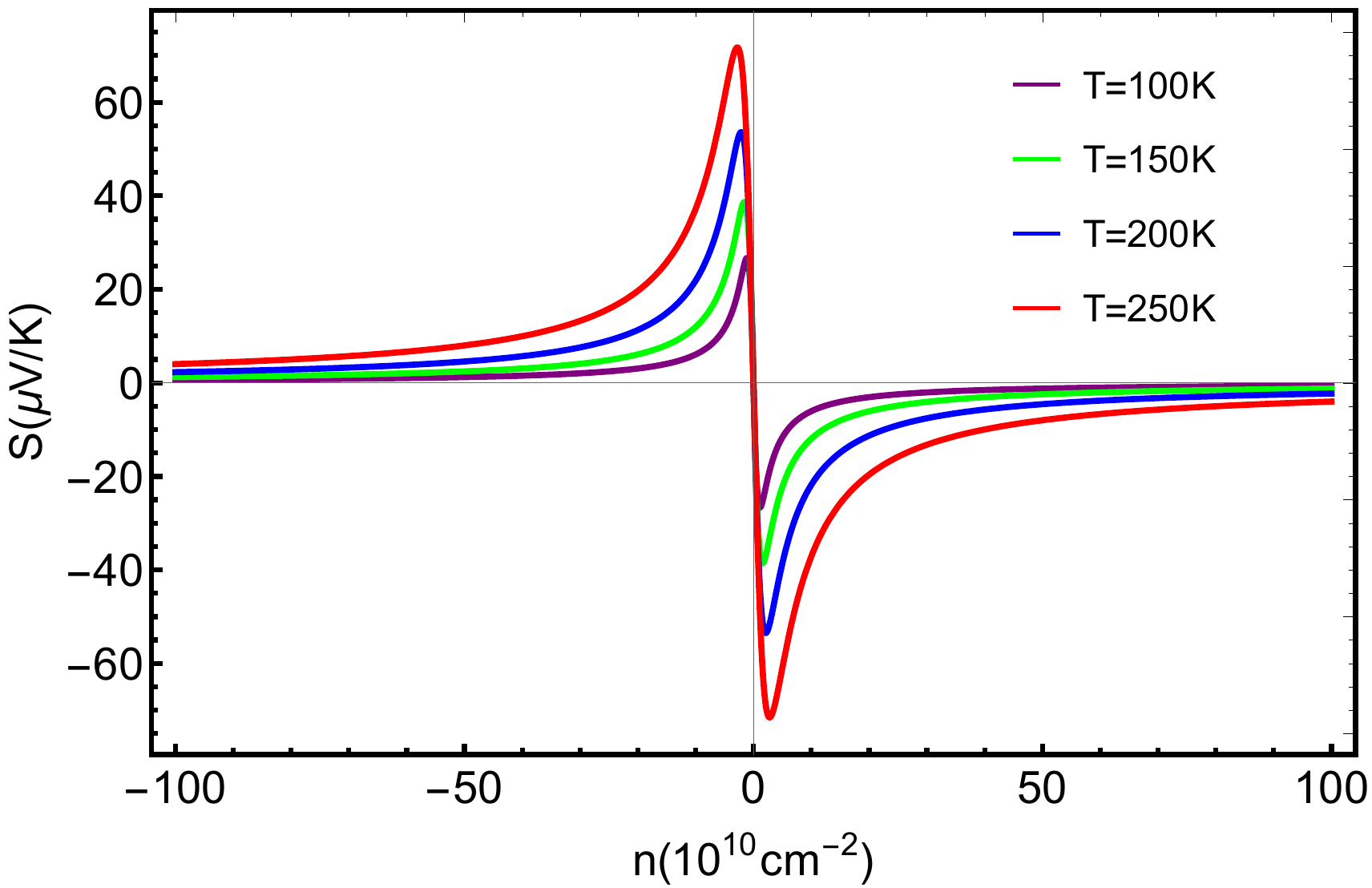}}
\subfigure{\includegraphics[width=7.6cm,height=5.9cm]{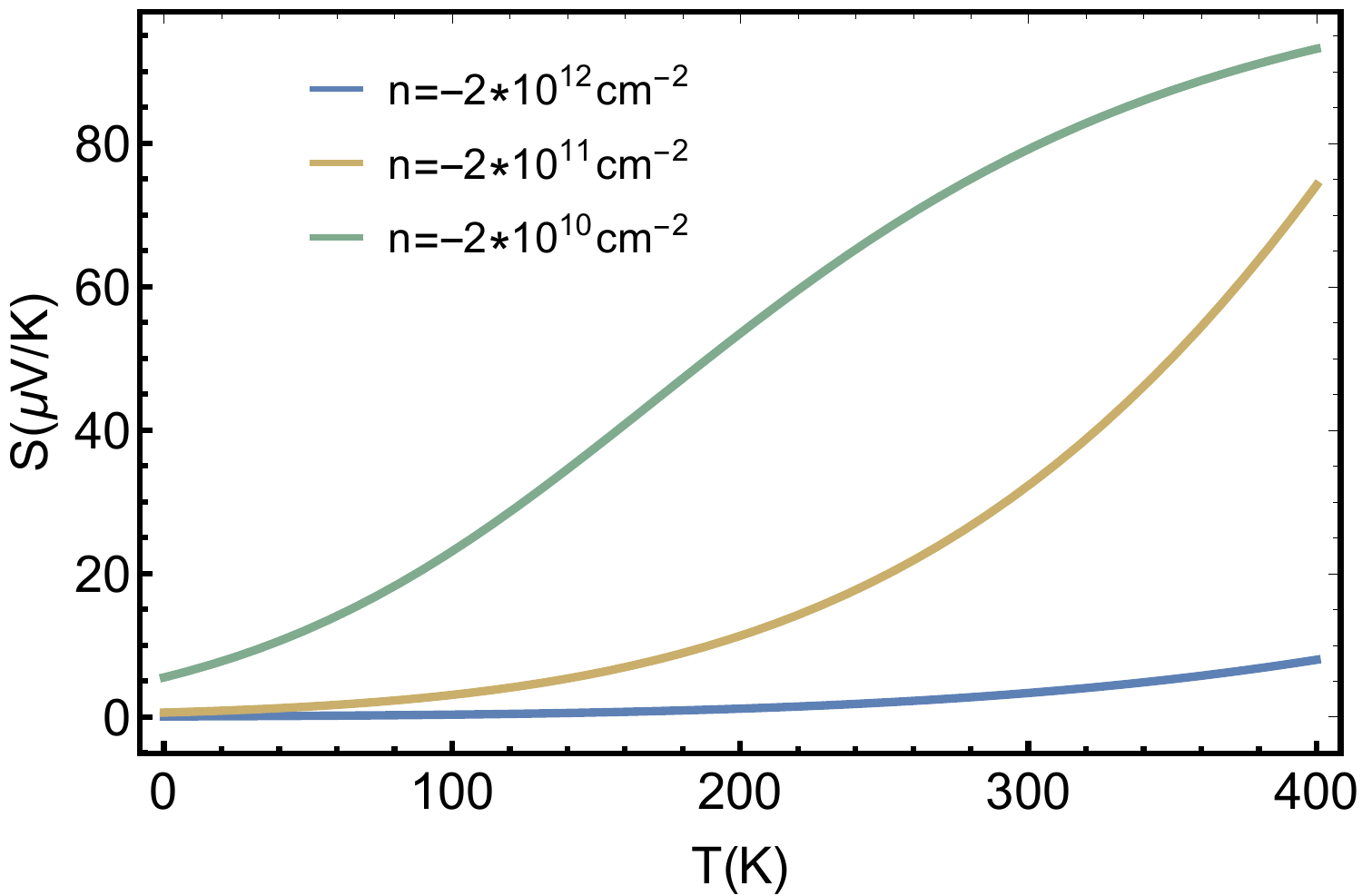}}
  \caption{\label{fig:alpha-see}(color online) (left panel) Measured Seebeck coefficient S as a function of density n for various temperatures. As we increase T, the maximum value of $S$ shifts and increases towards high concentration areas. (right panel) The measured Seebeck coefficient S as a function of temperatures for various density n. As we increase charge density n, the rate of change of S with temperature decreases, but it still maintains an upward trend.}
\end{figure}

Fig.\ref{alpha-T-n} and Fig.\ref{fig:alpha-see} show the relationship between the ther-
electric coefficient and the Seebeck coefficient as a function-
tion of temperature and carrier density. As the system crosses
the Fermi liquid region at high carrier densities, the enhanced-
ment decreases as disorder and scattering increase

Furthermore, near the neutrality point, the Seebeck coefficient increases with temperature. We also simulate the conductivity behavior with temperature in different carrier density regimes, showing that the rate of change of Seebeck coefficient with temperature is smaller near charge neutrality\cite{ghahari2016enhanced}. These results provide a valuable direction for future in-depth studies, as the variation of the Seebeck coefficient in graphene is influenced by multiple factors, including the transport properties of electrons and holes, the position of the Fermi level, and the temperature gradient.
  
\section{Parameter Impact on Transport Coefficients}

\begin{figure}[ht]
  \centering
\subfigure{\includegraphics[width=7.6cm,height=6cm]{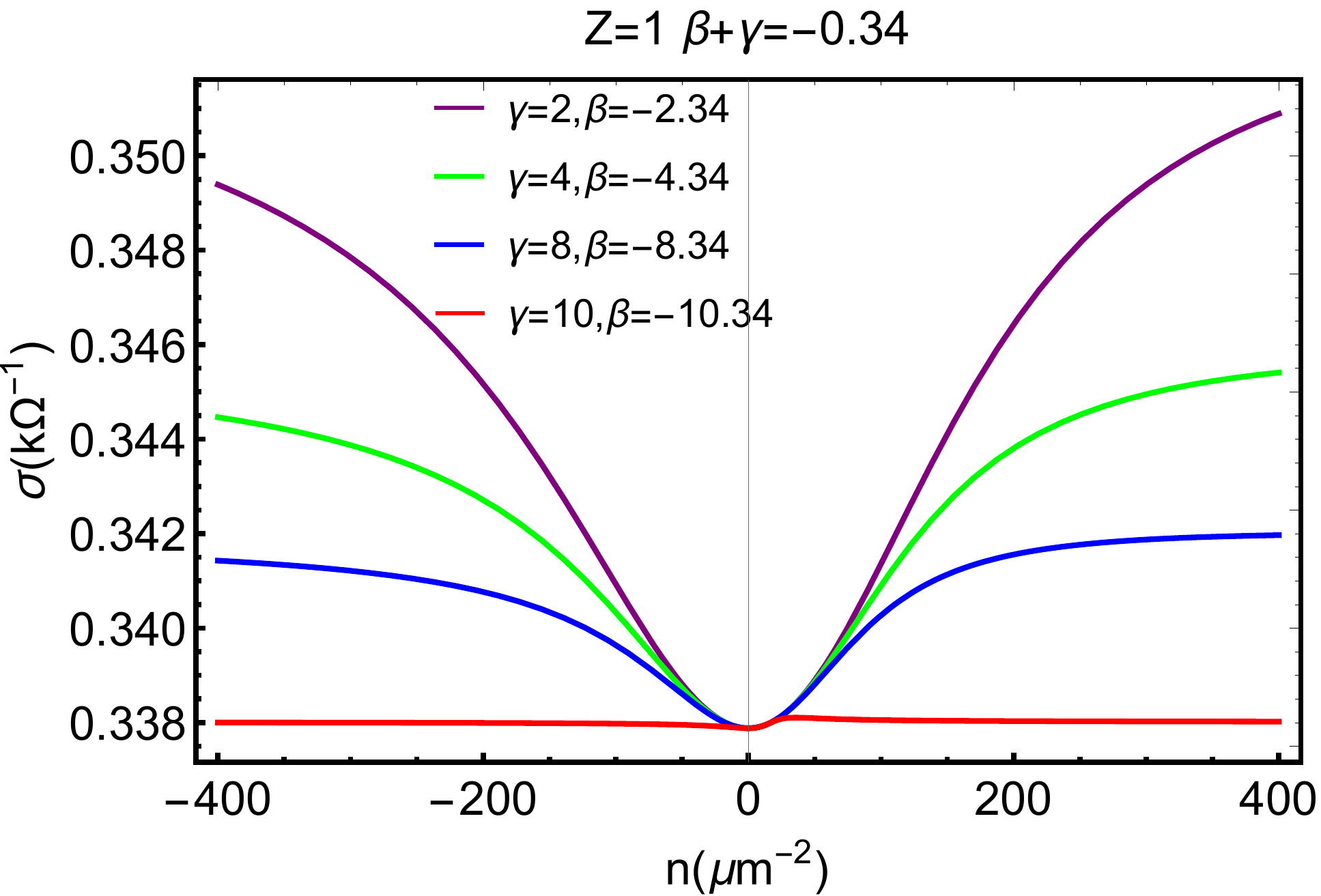}}
\subfigure{\includegraphics[width=7.6cm,height=5.9cm]{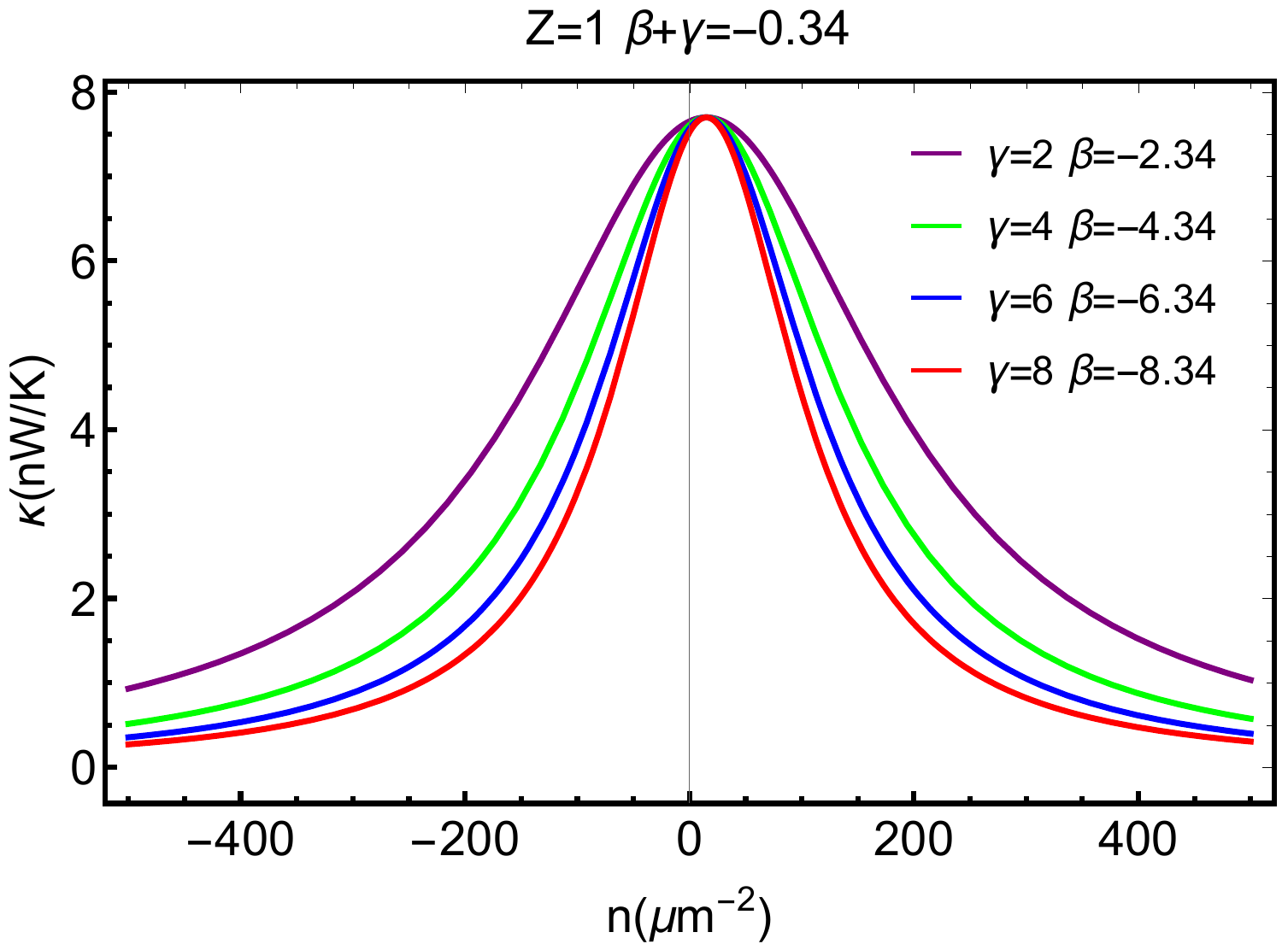}}
  \caption{\label{Parame}(color online) (left panel) DC electric conductivity $\sigma$ as a function of $n$ with different values of $\gamma$ and $\beta$ at $T=75K$. (right panel) DC thermal conductivity $\kappa$ as the function of n with different $\gamma$ and $\beta$ at $T=75K$. Due to the relationship between $\beta + \gamma$ and $Z$, it is necessary to ensure that their sum remains constant even when increasing $\beta$ and $\gamma$.}
\end{figure}

In this section we discuss the influence of the parameter $\beta+\gamma$ on $\sigma(n,T)$ and $\kappa(n,T)$. Under isothermal conditions, the electrical conductivity maintains a minimum near CNP. However, at higher densities, the increase of $\beta$ and $\gamma$ further suppresses $\sigma(n,T)$ and diminishes the characteristic features of graphene conductivity near the CNP, as shown in Fig.\ref{Parame}. This indicates a significant influence of $\beta$ and $\gamma$ parameters on the electrical conductivity under high density conditions.

According to $\sigma_{0}=0$ and $\kappa=\kappa_{0}$, the following relationships among $\beta$, $\gamma$, and $Z$ can be obtained:
\begin{equation}
\beta-\gamma=\frac{sZ}{g_{n}^2k^2\pi}\qquad 
\beta+\gamma=-\frac{sZ}{k^2\pi}
\end{equation}

At the same time, $\kappa(n,T)$ is also affected by these parameters, although to a lesser extent. While our model does not explicitly delineate the relative influence of $\beta$ and $\gamma$ on electrothermal transport, it reveals the potential for opposite motion of electrons and holes in an electric field. This hypothesis offers an intriguing perspective for interpreting the electrothermal transport properties of graphene, although more in-depth investigation is required to quantify the specific degree of influence.

\section{Summary and conclusion}
This article introduces the holographic dual current axion coupling model, which is first applied to the study of graphene Dirac fluids. By verifying the fluid properties at CNP and verifying the WF law at 75K, we found that the model is more accurate in matching experimental data compared to fluid mechanics methods. This study deepens our understanding of graphene Dirac fluids and also enhances the reliability of holographic duality methods.

Then we study transport properties, and observe the model aligns with experiments around $-100\mu m^{-2}\leq100\mu m^{-2}$, suggesting a potential entry into the Fermi liquid regime at higher concentrations. The definition of the crossover region between Dirac and Fermi fluids still requires precise determination.

In our model, $\sigma_0(n=0,T)$ incorporates an extra term $\beta+\gamma=-0.324$, accurately reflecting a linear increase in $\sigma_0$ with temperature. Our results are more precise compared to theories \cite{lucas2016transport,pongsangangan2022hydrodynamics} and \cite{seo2017holography,song2020determination}. We also explored the relationship between $\kappa_{0}$. While no decreasing trend in $\kappa(T)$was observed below 100K, our fitting better match that of Yunseok Seo et al.\cite{seo2017holography,song2020determination} The discussion about the WF law indicates that our theory shows that the WF law is violated in terms of temperature and charge dependence, indicating that graphene does not follow the properties of Fermi liquids near CNP. Our predicted results are consistent with those given by fluid mechanics, but there are still some discrepancies with experiments due to the thermal conductivity at high temperatures.

The model-based investigation delineates the dependence of the Seebeck coefficient S on carrier concentration at different temperatures. Results indicate an augmentation in the Seebeck coefficient with increasing temperature, with the maximum shifting towards higher carrier concentrations. The $\alpha$ coefficient diminishes as temperature rises. Discussion on the impact of $\beta+\gamma$ in electron and hole collisions reveals its inhibitory role in electrical conductivity, with a relatively minor effect on $\kappa(n,T)$. This suggests that these two parameters play a distinctive drag role in electron and hole scattering. 

Further scrutiny is needed to understand the influence of electron and phonon thermal transfer on thermal conductivity during electron and hole scattering processes at elevated temperatures. In the future, we anticipate holographic theories to provide a more accurate description of graphene's thermoelectric properties under high-temperature conditions.

\acknowledgements
We would like to thank Zhenhua Zhou and Jian-Pin Wu for their discussions and suggestions on improving our numerical algorithm.  Secondly, we also thank Yunseok Seo and Sang-Jin Sin for their valuable suggestions. 

\appendix
\section{Discussing the relationship between s, T, and n.}
\label{A1}

 From the temperature expression, it can be seen that when the temperature is 75K, the entropy density is still related to the charge density. Near the charge-neutral point, because our entropy density value is very large, the last term can be ignored and the entropy density can be considered as a function only of temperature, where$v_F=6*10^6m/s,g_n=2,n=100(\mu m)^{-2}$, and other parameters are obtained according to the values in the previous section, as shown below.
\begin{eqnarray}
&&T=\frac{\hbar v_F}{k_B}\frac{\sqrt{s}}{(4\pi)^{3/2}}(3-\frac{2\pi k^2}{s}-\frac{2\pi^2(1+g_n^2)n^2}{Z s^2})\\
&&s=\frac{3584.42}{(\mu m)^{-2}}
\end{eqnarray}
At the charge-neutral point $n=0$:
\begin{eqnarray}
&&T=\frac{\hbar v_F}{k_B}\frac{\sqrt{s}}{(4\pi)^{3/2}}(3-\frac{2\pi k^2}{s})\\
&&s=\frac{3538.33}{(\mu m)^{-2}}
\end{eqnarray}

Therefore, it can be seen that within the range of $n=100(\mu m)^{-2}$, our entropy density can be regarded as a function that only depends on temperature.

\section{Dimensional Analysis}
\label{A2}

We can restore units through dimensional analysis and set $\hbar=k_{B} =e=v_{F} =T=1 $, with the following settings before restoring the unit: $1/t\sim\omega,\hbar\omega=k_{B}T$. For thermal conductivity, there are:
\begin{equation}
\kappa=\frac{J^{Q}}{\nabla T}\sim\frac{\frac{\hbar \omega}{t}}{\frac{\hbar \omega}{k_{B}}}\sim\frac{\hbar \omega^2}{\frac{\hbar \omega}{k_{B}}}\sim
k_{B}\omega\sim\frac{k_{B}^2T}{\hbar}
\end{equation}

Therefore, the thermal conductivity at CNP before restoring the unit is: $\kappa_{0}=\frac {4\pi sT}{k^2}$, the thermal conductivity after reduction is:
\begin{equation}
\kappa_{0}^{*}=\frac{k_{B}^2 4\pi s T}{k^2\hbar}\sim\frac{k_{B}^2T}{\hbar}\sim \frac{k_{B}^2}{\hbar}\kappa_{0}
\end{equation}
For conductivity, there are:
\begin{eqnarray}
&&R=\frac{U}{I}=\frac{eU}{eI}\sim\frac{\hbar \omega}{\frac{e^2}{t}}\sim\frac{\hbar}{e^2}\\
&&\sigma=\frac{1}{R}\sim\frac{e^2}{\hbar}
\end{eqnarray}
Before reducing the conductivity of the unit CNP:
\begin{equation}
    \sigma_{0}=2(Z+\frac{k^2\pi(\beta+\gamma)}{s})
\end{equation}
After restoration, there will be:
\begin{equation}
\sigma_{0}^{*}=\frac{e^2}{\hbar}2(Z+\frac{k^2\pi(\beta+\gamma)}{s})
=\frac{e^2}{\hbar}\sigma_{0}
\end{equation}
The results of reducing the unit of conductivity and thermal conductivity when not at the neutral point of charge:
\begin{small}
\begin{eqnarray}
&&\kappa^{*}=\frac{\kappa_{0}^{*}}{1-\frac{2\pi Q^2 (g_{n}^2 k^2 \pi (\beta-\gamma)^{*})-sZ}{k^2 (s^2) (Z^2)}}\,,\\
&&\sigma^{*}=\sigma_{0}^{*}(\!1-\!\frac{n^2}{\frac{2\pi n^2k^2}{s\sigma_{0}}(\beta(\!1+\!g_n^2)+\!\gamma(\!1-\!g_n^2))\!-\!\frac{k^2 s Z^2}{\pi\sigma_{0}}})\,.
\end{eqnarray}
\end{small}
For the unit reduction of temperature, there are:
\begin{equation}
	T=\frac{\hbar v_F}{k_B}\frac{\sqrt{s}}{(4\pi)^{3/2}}(3-\frac{2\pi k^2}{s}-\frac{2\pi^2(1+g_n^2)n^2}{Z s^2})\,.
\end{equation}

\bibliographystyle{style1}
\bibliography{two-current}

\providecommand{\href}[2]{#2}\begingroup\raggedright\begin{thebibliography}{10}

\bibitem{crossno2016observation}
J.~Crossno, J.~K. Shi, K.~Wang, X.~Liu, A.~Harzheim, A.~Lucas, S.~Sachdev,
  P.~Kim, T.~Taniguchi, K.~Watanabe, et~al., {\it Observation of the dirac
  fluid and the breakdown of the wiedemann-franz law in graphene},  {\em
  Science} {\bf 351} (2016), no.~6277 1058--1061.

\bibitem{pongsangangan2022hydrodynamics}
K.~Pongsangangan, T.~Ludwig, H.~T. Stoof, and L.~Fritz, {\it Hydrodynamics of
  charged dirac electrons in two dimensions. ii. role of collective modes},
  {\em Phys. Rev. B} {\bf 106} (2022), no.~20 205127.

\bibitem{foster2009slow}
M.~S. Foster and I.~L. Aleiner, {\it Slow imbalance relaxation and
  thermoelectric transport in graphene},  {\em Phys. Rev. B} {\bf 79} (2009),
  no.~8 085415.

\bibitem{alishahiha2012charged}
M.~Alishahiha, E.~{\'O}. Colg{\'a}in, and H.~Yavartanoo, {\it Charged black
  branes with hyperscaling violating factor},  {\em J. High Energy Phys.} {\bf
  2012} (2012), no.~11 1--20.

\bibitem{amoretti2014thermo}
A.~Amoretti, A.~Braggio, N.~Maggiore, N.~Magnoli, and D.~Musso, {\it
  Thermo-electric transport in gauge/gravity models with momentum dissipation},
   {\em J. High Energy Phys.} {\bf 2014} (2014), no.~9 1--30.

\bibitem{amoretti2015analytic}
A.~Amoretti, A.~Braggio, N.~Maggiore, N.~Magnoli, and D.~Musso, {\it Analytic
  dc thermoelectric conductivities in holography with massive gravitons},  {\em
  Phys. Rev. D} {\bf 91} (2015), no.~2 025002.

\bibitem{banks2015thermoelectric}
E.~Banks, A.~Donos, and J.~P. Gauntlett, {\it Thermoelectric dc conductivities
  and stokes flows on black hole horizons},  {\em J. High Energy Phys.} {\bf
  2015} (2015), no.~10 1--38.

\bibitem{cheng2015thermoelectric}
L.~Cheng, X.-H. Ge, and Z.-Y. Sun, {\it Thermoelectric dc conductivities with
  momentum dissipation from higher derivative gravity},  {\em J. High Energy
  Phys.} {\bf 2015} (2015), no.~4 1--16.

\bibitem{donos2015navier}
A.~Donos and J.~P. Gauntlett, {\it Navier-stokes equations on black hole
  horizons and dc thermoelectric conductivity},  {\em Phys. Rev. D} {\bf 92}
  (2015), no.~12 121901.

\bibitem{donos2016dc}
A.~Donos, J.~P. Gauntlett, T.~Griffin, and L.~Melgar, {\it Dc conductivity of
  magnetised holographic matter},  {\em J. High Energy Phys.} {\bf 2016}
  (2016), no.~1 1--37.

\bibitem{ling2016characterization}
Y.~Ling, P.~Liu, and J.-P. Wu, {\it Characterization of quantum phase
  transition using holographic entanglement entropy},  {\em Phys. Rev. D} {\bf
  93} (2016), no.~12 126004.

\bibitem{ling2016novel}
Y.~Ling, P.~Liu, and J.-P. Wu, {\it A novel insulator by holographic
  q-lattices},  {\em J. High Energy Phys.} {\bf 2016} (2016), no.~2 1--12.

\bibitem{ling2018holographic}
Y.~Ling, P.~Liu, J.-P. Wu, and M.-H. Wu, {\it Holographic superconductor on a
  novel insulator},  {\em Chin. Phys. C} {\bf 42} (2018), no.~1 013106.

\bibitem{ling2021instability}
Y.~Ling and M.-H. Wu, {\it Instability of ads black holes with lattices},  {\em
  Chin. Phys. C} {\bf 45} (2021), no.~2 025102.

\bibitem{seo2017holography}
Y.~Seo, G.~Song, P.~Kim, S.~Sachdev, and S.-J. Sin, {\it Holography of the
  dirac fluid in graphene with two currents},  {\em Phys. Rev. Lett.} {\bf 118}
  (2017), no.~3 036601.

\bibitem{gouteraux2016effective}
B.~Gout{\'e}raux, E.~Kiritsis, and W.-J. Li, {\it Effective holographic
  theories of momentum relaxation and violation of conductivity bound},  {\em
  J. High Energy Phys.} {\bf 2016} (2016), no.~4 1--23.

\bibitem{baggioli2017higher}
M.~Baggioli, B.~Gout{\'e}raux, E.~Kiritsis, and W.-J. Li, {\it Higher
  derivative corrections to incoherent metallic transport in holography},  {\em
  J. High Energy Phys.} {\bf 2017} (2017), no.~3 1--32.

\bibitem{li2019simple}
W.-J. Li and J.-P. Wu, {\it A simple holographic model for spontaneous breaking
  of translational symmetry},  {\em Eur. Phys. J. C} {\bf 79} (2019), no.~3
  1--8.

\bibitem{liu2022alternating}
Y.~Liu, X.-J. Wang, J.-P. Wu, and X.~Zhang, {\it Alternating current
  conductivity and superconducting properties of the holographic effective
  theory},  {\em arXiv preprint arXiv:2201.06065} (2022).

\bibitem{zhong2022transverse}
Y.-Y. Zhong and W.-J. Li, {\it Transverse goldstone mode in holographic fluids
  with broken translations},  {\em Eur. Phys. J. C} {\bf 82} (2022), no.~6 511.

\bibitem{donos2014novel}
A.~Donos and J.~P. Gauntlett, {\it Novel metals and insulators from
  holography},  {\em J. High Energy Phys.} {\bf 2014} (2014), no.~6 1--26.

\bibitem{ling2017holographic}
Y.~Ling, P.~Liu, J.-P. Wu, and Z.~Zhou, {\it Holographic metal-insulator
  transition in higher derivative gravity},  {\em Phys. Lett. B} {\bf 766}
  (2017) 41--48.

\bibitem{baggioli2017diffusivities}
M.~Baggioli and W.-J. Li, {\it Diffusivities bounds and chaos in holographic
  horndeski theories},  {\em J. High Energy Phys.} {\bf 2017} (2017), no.~7
  1--26.

\bibitem{2014JHEP...11..081D}
A.~{Donos} and J.~P. {Gauntlett}, {\it {Thermoelectric DC conductivities from
  black hole horizons}},  {\em J. High Energy Phys.} {\bf 2014} (Nov., 2014)
  81, [\href{http://arxiv.org/abs/1406.4742}{{\tt arXiv:1406.4742}}].

\bibitem{Blake2013UniversalRF}
M.~Blake and D.~Tong, {\it Universal resistivity from holographic massive
  gravity},  {\em Phys. Rev. D} {\bf 88} (2013) 106004.

\bibitem{blake2015quantum}
M.~Blake and A.~Donos, {\it Quantum critical transport and the hall angle in
  holographic models},  {\em Phys. Rev. Lett} {\bf 114} (2015), no.~2 021601.

\bibitem{sarma2011electronic}
S.~D. Sarma, S.~Adam, E.~Hwang, and E.~Rossi, {\it Electronic transport in
  two-dimensional graphene},  {\em Rev. Mod. Phys.} {\bf 83} (2011), no.~2 407.

\bibitem{basov2014colloquium}
D.~Basov, M.~Fogler, A.~Lanzara, F.~Wang, Y.~Zhang, et~al., {\it Colloquium:
  graphene spectroscopy},  {\em Rev. Mod. Phys.} {\bf 86} (2014), no.~3 959.

\bibitem{hartnoll2007theory}
S.~A. Hartnoll, P.~K. Kovtun, M.~M{\"u}ller, and S.~Sachdev, {\it Theory of the
  nernst effect near quantum phase transitions in condensed matter and in
  dyonic black holes},  {\em Phys. Rev. B} {\bf 76} (2007), no.~14 144502.

\bibitem{lucas2018hydrodynamics}
A.~Lucas and K.~C. Fong, {\it Hydrodynamics of electrons in graphene},  {\em J.
  Phys. : Condens.Matter} {\bf 30} (2018), no.~5 053001.

\bibitem{li2022transport}
Y.-L. Li, X.-J. Wang, G.~Fu, and J.-P. Wu, {\it Transport properties of a
  3-dimensional holographic effective theory with gauge-axion coupling},  {\em
  Phys. Lett. B} {\bf 829} (2022) 137124.

\bibitem{iqbal2009universality}
N.~Iqbal and H.~Liu, {\it Universality of the hydrodynamic limit in ads/cft and
  the membrane paradigm},  {\em Phys. Rev. D} {\bf 79} (2009), no.~2 025023.

\bibitem{donos2014thermoelectric}
A.~Donos and J.~P. Gauntlett, {\it Thermoelectric dc conductivities from black
  hole horizons},  {\em J. High Energy Phys.} {\bf 2014} (2014), no.~11 1--25.

\bibitem{kim2015thermoelectric}
K.-Y. Kim, K.~K. Kim, Y.~Seo, and S.-J. Sin, {\it Thermoelectric conductivities
  at finite magnetic field and the nernst effect},  {\em J. High Energy Phys.}
  {\bf 2015} (2015), no.~7 1--29.

\bibitem{lucas2016transport}
A.~Lucas, J.~Crossno, K.~C. Fong, P.~Kim, and S.~Sachdev, {\it Transport in
  inhomogeneous quantum critical fluids and in the dirac fluid in graphene},
  {\em Phys. Lett. B} {\bf 93} (2016), no.~7 075426.

\bibitem{song2020determination}
G.~Song, Y.~Seo, and S.-J. Sin, {\it Determination of dynamical exponents of
  graphene at quantum critical point by holography},  {\em Phys. Rev. D} {\bf
  102} (2020), no.~12 126023.

\bibitem{narozhny2021hydrodynamic}
B.~N. Narozhny and I.~V. Gornyi, {\it Hydrodynamic approach to electronic
  transport in graphene: energy relaxation},  {\em Front. Phys} {\bf 9} (2021)
  640649.

\bibitem{ziman2001electrons}
J.~M. Ziman, {\em Electrons and phonons: the theory of transport phenomena in
  solids}.
\newblock Oxford university press, 2001.

\bibitem{mermin1978solid}
A.~Mermin, {\it Solid state physics. von nw ashcroft und nd mermin; holt,
  rinehart and winston, new york 1976, xxii, 826 seiten,$19,95$},  {\em Phys.
  unserer Zeit} {\bf 9} (1978), no.~1 33--33.

\bibitem{mahajan2013non}
R.~Mahajan, M.~Barkeshli, and S.~A. Hartnoll, {\it Non-fermi liquids and the
  wiedemann-franz law},  {\em Phys. Rev. B} {\bf 88} (2013), no.~12 125107.

\bibitem{kim2009violation}
K.-S. Kim and C.~P{\'e}pin, {\it Violation of the wiedemann-franz law at the
  kondo breakdown quantum critical point},  {\em Phys. Rev. Lett} {\bf 102}
  (2009), no.~15 156404.

\bibitem{fong2012ultrasensitive}
K.~C. Fong and K.~Schwab, {\it Ultrasensitive and wide-bandwidth thermal
  measurements of graphene at low temperatures},  {\em Phys. Rev. X} {\bf 2}
  (2012), no.~3 031006.

\bibitem{fong2013measurement}
K.~C. Fong, E.~E. Wollman, H.~Ravi, W.~Chen, A.~A. Clerk, M.~Shaw, H.~Leduc,
  and K.~Schwab, {\it Measurement of the electronic thermal conductance
  channels and heat capacity of graphene at low temperature},  {\em Phys. Rev.
  X} {\bf 3} (2013), no.~4 041008.

\bibitem{crossno2015development}
J.~Crossno, X.~Liu, T.~A. Ohki, P.~Kim, and K.~C. Fong, {\it Development of
  high frequency and wide bandwidth johnson noise thermometry},  {\em Appl.
  Phys. Lett.} {\bf 106} (2015), no.~2.

\bibitem{ghahari2016enhanced}
F.~Ghahari, H.-Y. Xie, T.~Taniguchi, K.~Watanabe, M.~S. Foster, and P.~Kim,
  {\it Enhanced thermoelectric power in graphene: Violation of the mott
  relation by inelastic scattering},  {\em Phys. Rev. Lett} {\bf 116} (2016),
  no.~13 136802.

\end{thebibliography}\endgroup
\end{document}